\documentclass[
 reprint,prl,
 superscriptaddress,
 amsmath,amssymb,
 aps
]{revtex4-2}

\usepackage[utf8]{inputenc}
\usepackage{mathtools}
\usepackage{amsmath,amssymb}
\usepackage{xcolor}
\usepackage{graphicx}
\usepackage{soul}
\usepackage{cancel}
\usepackage{fullpage}
\usepackage{mdframed}
\definecolor{mycol}{RGB}{0,102,204} 
\usepackage[colorlinks, linkcolor=mycol, citecolor=mycol, urlcolor=black, breaklinks]{hyperref}
\usepackage[capitalise]{cleveref}

\newcommand{\im}{{\rm Im\,}}
\newcommand{\re}{{\rm Re\,}}

\newcommand{\tr}{{\rm Tr}}

\newcommand{\dint}{{\int\hskip -4mm\diagup \,\,}}

\providecommand{\ketbra}[1]{|#1\rangle\!\langle#1|}
\providecommand{\ket}[1]{|#1\rangle}

\newcommand{\red}[1]{{\color{blue}{#1}}}

\newif\ifincludesm
\includesmtrue  

\newif\ifmblastbibitemdone

\ifincludesm
\usepackage[resetlabels]{multibib}
\newcites{SM}{Supplemental References}
\usepackage{etoolbox}
\makeatletter
\AtBeginDocument{%
  \apptocmd{\std@thebibliography}{%
    \let\@TBN@opr\present@bibnote
    \@FMN@list
    \global\let\@FMN@list\@empty
  }{}{}%
  \pretocmd{\endthebibliography}{%
    \ifmblastbibitemdone\else
      \label{LastBibItem}\global\mblastbibitemdonetrue
    \fi
  }{}{}%
}
\let\mb@sm@bibliography\bibliographySM
\renewcommand\bibliographySM[1]{%
  \begingroup
    \def\@biblabel##1{[S##1]}%
    \mb@sm@bibliography{#1}%
  \endgroup}
\makeatother
\fi

\begin{document}

\title{Universal dynamics from a single-particle dark state}

\author{Ruben Daraban}
\affiliation{CESQ/ISIS (UMR 7006), CNRS and Universit\'{e} de Strasbourg, 67000 Strasbourg, France}
\author{Arghavan Safavi-Naini}
\affiliation{Institute for Theoretical Physics, Institute of Physics, University of Amsterdam, Science Park 904, 1098 XH Amsterdam, The Netherlands}
\affiliation{QuSoft, Science Park 123, 1098 XG Amsterdam, The Netherlands}

\author{Johannes Schachenmayer}
\affiliation{CESQ/ISIS (UMR 7006), CNRS and Universit\'{e} de Strasbourg, 67000 Strasbourg, France}

\author{Mohammad Maghrebi}
\email{maghrebi@msu.edu}
\affiliation{Department of Physics and Astronomy, Michigan State University, East Lansing, Michigan 48824 USA}

\begin{abstract}
Open quantum systems can host dark or subradiant states whose decay is highly suppressed. While these states have been extensively studied in the few-excitation regime, their impact on the many-body dynamics remains largely unexplored. Here, we study a spin chain  subject to correlated dissipation on neighboring sites, which admits a single-particle dark state at zero momentum. We show that the single-particle dark state qualitatively alters the many-body dynamics at long times, and identify its distinct universal behavior. While the zero-momentum mode is dark at the single-particle level, it decays slowly as $1/\log t$ as it becomes dressed by other modes through a dissipation-induced nonlinearity.   We demonstrate that the momentum distribution takes a universal scaling form in $k\sqrt{t}$, and the total density decays as $1/\sqrt{t}\log t$. Our results further elucidate the origin of the conflicting results in recent works. Finally, we corroborate the analytics with matrix product state simulations and show that the same universal behavior persists for soft-core interactions, underscoring the universality of the emergent dynamics.
\end{abstract}

\maketitle

Many-body open quantum systems have attracted much attention in the last two decades. On one hand, they are very relevant to existing quantum simulators which emulate many-body physics of interest but also unavoidably come with noise and decoherence \cite{preskill_quantum_2018}. On the other hand, tailored dissipation can be a resource itself, for example, for quantum state preparation \cite{diehl_quantum_2008,verstraete_quantum_2009,harrington_engineered_2022}. Furthermore, dissipation competes with Hamiltonian dynamics in driven open quantum systems, giving rise to genuinely novel forms of non-equilibrium physics \cite{sieberer_universality_2025}. 

Dissipative systems can host subradiant or \textit{dark} states that are long-lived or do not decay. 
Subradiance arises when  photons emitted by atoms interfere destructively suppressing their decay \cite{prasad_polarium_2000,svidzinsky_cooperative_2010}. Integrating out photons then leads to  correlated long-range atomic dissipation, resulting in momentum-dependent dissipation which could be highly suppressed for subradiant states, a scenario that naturally occurs in waveguide QED \cite{sheremet_waveguide_2023}. 
While most studies of subradiance have focused on states with only a few atomic excitations \cite{asenjogarcia_exponential_2017, guerin_subradiance_2016,sutherland_collective_2016,bettles_cooperative_2016,jen_cooperative_2016,kornovan_collective_2016,haakh_polaritonic_2016,ruostekoski_emergence_2016,ruostekoski_arrays_2017,santo_collective_2020}, the many-body regime, with a few exceptions \cite{albrecht_subradiant_2019,masson_many_2020}, remains largely unexplored. 
Interestingly, states with few excitations are shown to exhibit a kind of \textit{fermionization} \cite{asenjogarcia_exponential_2017,albrecht_subradiant_2019,zhang_theory_2019,zhang_subradiant_2020_a,zhang_free_2022}, which also leaves a distinct fingerprint on the long-time dynamics \cite{henriet_critical_2019}. 
In fact, even a single-particle dark state can drastically affect many-body dynamics since it results in a slow mode. 
A minimal such model, in which a spin chain is subject to correlated dissipation on neighboring sites, has recently attracted considerable attention \cite{begg_quantum_2024,pocklington_efficient_2025,marche_open_2026}. This model admits a single-particle dark state, which leads to slow dynamics at long times. 
However, different works report conflicting results and even different critical exponents. 
At a technical level, while the Hamiltonian itself can be mapped to free fermions, dissipation renders the model highly nonlinear \cite{pocklington_efficient_2025,marche_open_2026}.  

\begin{figure}[b!]
    \centering
    \includegraphics[width=0.85\linewidth]{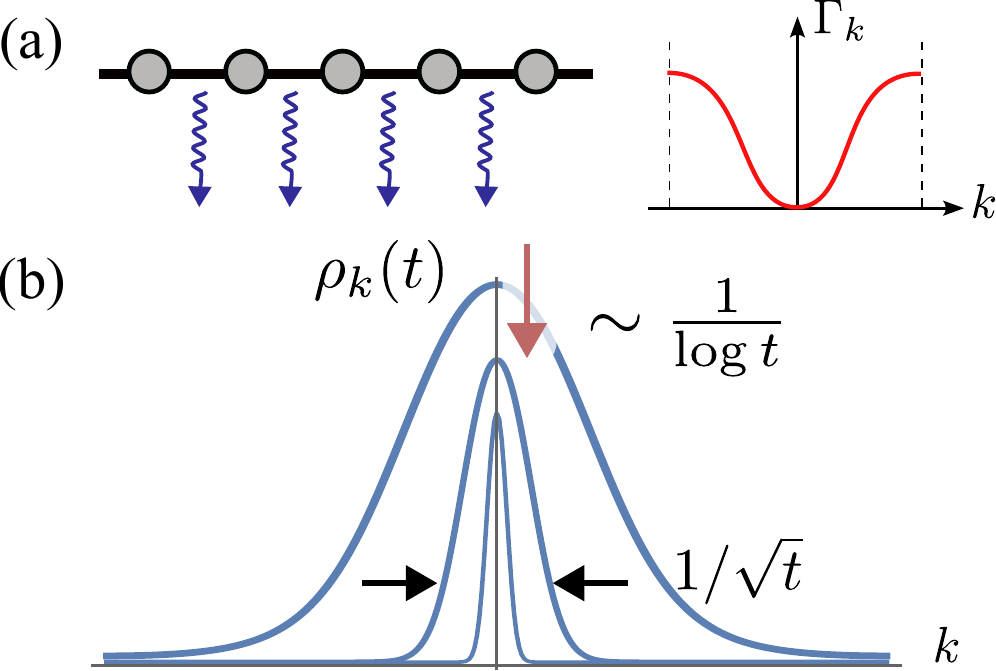}
    \caption{(a) Schematic representation of a spin chain (alternatively, hard- or soft-core bosons) subject to correlated dissipation on neighboring sites featuring a single-particle dark state at zero momentum ($\Gamma_{k=0}=0$). (b)~The momentum distribution at long times exhibits a diffusion-like behavior with the width decaying as $\sim 1/\sqrt{t}$, as if governed by 
    the single-particle decay rate $\Gamma_k \sim k^2$. However, the $k=0$ mode also decays, albeit slowly, as it becomes dressed with other modes due to a dissipation-induced nonlinearity. This decay takes a universal form of an inverse logarithm.} 
    \label{fig:schematic}
\end{figure}

In this work, we characterize the exact nature of the universal dynamics in the presence of a single-particle dark state (the  $k=0$ mode). 
We find that this mode still decays due to the nonlinear coupling to other modes which are dissipative themselves. At long times, this decay takes the universal form of $1/\log t$; see \cref{fig:schematic}. Furthermore, we show that the momentum distribution takes a universal scaling form (in $k\sqrt{t}$), and that the total density decays as $n(t)\propto 1/(\sqrt{t}\log t)$, and further elucidate the origin of the conflicting results in the literature. We obtain these results analytically at weak dissipation using an ansatz of free fermions (i.e., a Gaussian state) whose characteristics (i.e., generalized temperatures) evolve nonlinearly due to dissipation.
We corroborate the analytics with numerical results using matrix product states (MPS), and further show that soft-core interactions exhibit the same behavior, underscoring the universality of our results.

\textit{Model.---}Here, we consider an open quantum spin chain whose dynamics is governed by the quantum master equation
\begin{equation}
    \frac{{\rm d}\hat \rho}{{\rm d}t}=-i[\hat H,\hat \rho]+\Gamma \sum_j \hat L_j \hat \rho \hat L_j^\dagger-\frac{1}{2}\{\hat L^\dagger_j \hat L_j,\hat \rho\}\,,
\end{equation}
where $\hat H$ is the XX Hamiltonian 
\begin{align}\label{eq:Hamilt}
    \hat H 
    = -J\sum_j
    \sigma^x_j \sigma^x_{j+1}+ \sigma^y_j \sigma^y_{j+1}\,,
\end{align}
and the Lindblad operator is given by  
\begin{equation}
    \hat L_j= \sigma^-_{j+1}- \sigma^-_{j}\,,
\end{equation}
describing a correlated type of dissipation; see \cref{fig:schematic}. This type of dissipation appears in waveguide QED \cite{sheremet_waveguide_2023}, trapped ions \cite{begg_quantum_2024}, Rydberg systems facilitated by the EIT \cite{peyronel_quantum_2012,baldwin_singularities_2021}, as well as superconducting circuits \cite{marino_driven_2016,kamar_markovian_2025}. 
A trivial steady state of the dynamics is the state with all spins down.
Dissipation also admits a dark state formed by a uniform superposition of single spin excitations. This becomes evident by writing the Lindblad operator in momentum space, $\hat L_k =(e^{ik}-1)\sigma_k^-$, indicating that spin wave excitations decay at the rate $\Gamma_k = \Gamma |e^{ik}-1|^2$, resulting in a zero mode at $k=0$. On the other hand, a finite density of spin-wave excitations interact in a nontrivial fashion due to the hard-core nature of spin~$\frac{1}{2}$, likely inducing the zero mode to decay. While the steady state is trivial, the slow dynamics of the long-wavelength modes remains  highly nontrivial.

\textit{Mapping to fermions.---}To tackle the dynamics, we first note that the Hamiltonian can be exactly mapped to free fermions via the Jordan-Wigner transformation, $\hat H= \sum_k \omega_k \hat c_k^\dagger \hat c_k$ with $\hat c_k (\hat c^\dagger_k)$ annihilation (creation) operators. However, dissipation spoils integrability since (infinitely) long string operators appear due to the jump term ($\hat L_j\hat \rho \hat L_j^\dagger$). 
If we naively drop the strings, the Lindblad operator takes the form $\hat L_k=\sqrt{\Gamma_k} \hat c_k$. At long times, we find the fermionic density (or, the \textit{rapidity}), $\rho_k\equiv \langle \hat c_k^\dagger \hat c_k\rangle$, decays as $\rho_k(t) \propto e^{-\Gamma k^2 t}$. In this approximation, $\rho_{k=0}$ does not decay while the total density, $n=\int \frac{{\rm d}k}{2\pi} \rho_k$, decays as $n(t)\sim 1/\sqrt{t}$.  But as we argued before, we expect the $k=0$ mode to decay due to the coupling to other modes, a feature that is not captured by free fermions. 

More generally, one can still find an exact equation but only for the decay of the total density:
\begin{equation}\label{eq:n_from_m}
    \dot n= \int \frac{{\rm d}k}{2\pi} \dot \rho_k = -\int \frac{{\rm d}k}{2\pi} \Gamma_k \rho_k \equiv -\Gamma m(t)\,,
\end{equation}
where we have defined $m$ for later convenience. 
This equation follows from the Heisenberg equation for the number operator and the fact that the Hamiltonian is number conserving \cite{sm}. 
Now, comparing the integrals in the above equation and using $\Gamma_k \sim k^2$ at long wavelengths suggests a scaling solution for $\rho_k(t)$ as a function of $k\sqrt{\Gamma t}$. While the free-fermion picture satisfies this scaling, a complete solution should have some form of nonlinearity.  

\textit{Perturbative approach.---}To go beyond free fermions, we take a perturbative approach where dissipation is assumed weak compared to the unitary dynamics, $\Gamma\ll J$. In this limit, we can assume the system takes the form of a generalized Gibbs ansatz constructed from fermionic populations $\hat c_k^\dagger \hat c_k$ where the corresponding generalized effective temperatures become time-dependent \cite{lange_pumping_2017,lange_time_2018}. Employing this ansatz in the quantum master equation gives the dynamics of these temperatures or equivalently those of $\rho_k(t)$. Parity considerations complicate this picture as the fermionic parity changes when a quantum jump occurs. A general analysis properly treating the two parity sectors leads to nonlinear dynamics for $\rho_k$, and has been recently applied in several contexts \cite{bouchoule_effect_2020,riggio_effects_2024,ali_signatures_2026}. The resulting equation is a type of kinetic equation, 
\begin{subequations}\label{eq:dr/dt}
\begin{equation}
\frac{1}{\Gamma}\frac{\rm d \rho_k }{{\rm d} t} =   -F_k(\{\rho_p\})\,,
\end{equation}
where $F_k$ is a nonlinear function of the $\rho_p$s. For the model considered in this work, we find
\begin{align}\label{eq:F_rho_k}
 \begin{split}
 F_k
 = &|\delta_k|^2\rho_k(1-\rho_k) +\left|\dint \frac{{\rm d}p}{\pi} \frac{\delta_p\rho_p}{\delta_{p-k}}\right|^2\\
&+2m \dint \frac{{\rm d}p}{\pi} \frac{\rho_k-\rho_p}{|\delta_{p-k}|^2}\,,
\end{split}
\end{align}
\end{subequations}
where $\delta_k\equiv 1-e^{ik}$, and the slash denotes the Cauchy principal value integral. The resulting equation is thus
a nonlinear integro-differential equation. The linear term on the rhs of this equation is just that of free fermions; however, the nonlinear terms should be considered on equal footing. This equation becomes exact in the limit $\Gamma\to 0$ for any finite $\Gamma t$.  
We leave the derivation of the above equation to the SM \cite{sm}. This equation was recently derived using a different (and somewhat simpler) treatment assuming open boundary conditions and carefully analyzed numerically \cite{marche_open_2026}. However, the nature of the universal dynamics has remained unresolved. Here, we provide sharp analytical insights together with large-scale MPS simulations.

\textit{Decay of the zero mode.---}Next, we specialize the nonlinear equation for $\rho_k$ to the $k=0$ mode:
\begin{equation}\label{eq:rho_0}
    \frac{1}{\Gamma}\frac{\rm d \rho_0 }{{\rm d} t}=-4n(t)^2-2m(t)\int \frac{{\rm d}p}{\pi} \frac{\rho_0(t)-\rho_p(t)}{|1-e^{ip}|^2}\,.
\end{equation}
The expression on the rhs is purely nonlinear in $\rho_k$ (recall the definition of $n(t), m(t)$), and would be otherwise zero for free fermions. 
The  first term can be interpreted as the loss due to the nonlinearity proportional to the square of the density. The second term, also a nonlinear contribution, reshuffles the population among different momenta (unlike the Hamiltonian). This term is also negative, leading to further loss of $\rho_0$, if the density at $k=0$ is higher than that at nonzero $k$. 
To characterize the dynamics, we postulate a scaling solution at long times as
\begin{equation}\label{eq:scaling_rho_k}
    \rho_k(t)=\rho_0(t) f(k\sqrt{\Gamma t})\,,
\end{equation}
where $f$ is a scaling function with the properties that $f(0)=1$ and $\lim_{x\to \infty}f(x)= 0$; again, the rescaled variable $k\sqrt{\Gamma t}$ is motivated by \cref{eq:n_from_m}. 
Plugging this scaling solution in \cref{eq:rho_0} (also replacing $1/|1-e^{ip}|^2 \approx 1/p^2$ in the spirit of a scaling solution) together with \cref{eq:n_from_m}, we find an exact solution for $\rho_0(t)$. A simple form of the solution emerges at long times as
\begin{equation}\label{eq:rho0_log}
    \rho_0(t) \sim \frac{1}{\rho_0(\tau)^{-1} + c \log \Gamma t}\,,
\end{equation}
where $c$ can be computed from integrals of the scaling function $f$ and thus is a universal number. Specifically,  for $f(x)=e^{- x^2}$ describing free fermions, we find $c=2/\pi$. This is indeed consistent with the plots in \cref{fig:rho_from_ODE} obtained from a full numerical solution of \cref{eq:dr/dt}. We also provide an analytical argument shortly. 
Finally, $\rho_0(\tau)$ indicates the value of the $k=0$ density at time $\tau\sim 1/\Gamma$ when the scaling solution sets in. 

\Cref{eq:scaling_rho_k,eq:rho0_log} immediately lead to $n(t)\sim 1/\sqrt{t}\log t$ at long times with a universal multiplicative logarithm. The different critical exponents identified in the literature \cite{begg_quantum_2024,pocklington_efficient_2025,marche_open_2026} are thus partially explained by the fact that logarithms may appear as small exponents. What makes resolving the decay even more difficult is that the constant term in the denominator in \cref{eq:rho0_log} can be rather large, thus effectively masking the logarithm for long times. 
To see this, 
we can make an approximation that, up to the time $\tau$, the density $\rho_k(t)$ is roughly uniform in $k$. 
\Cref{eq:rho_0} then becomes $\dot \rho_0 \sim -4\Gamma\rho_0^2$, leading to the solution $\rho_0(t) =1/(\rho_0(0)^{-1}+ 4 \Gamma t)$.
This rough solution works rather well at short times \cite{sm}, leading to a fast initial decay. 
For a fully excited initial state ($\rho_0(0)=1$) the logarithm in \cref{eq:rho0_log} only dominates the constant for times $\Gamma t \gtrsim 10^4$. The many decades of dynamics required underscore the difficulty in identifying the universal nature of the dynamics in the numerics. 

\begin{figure}[t!]
    \centering
\includegraphics[width=1.0\linewidth]{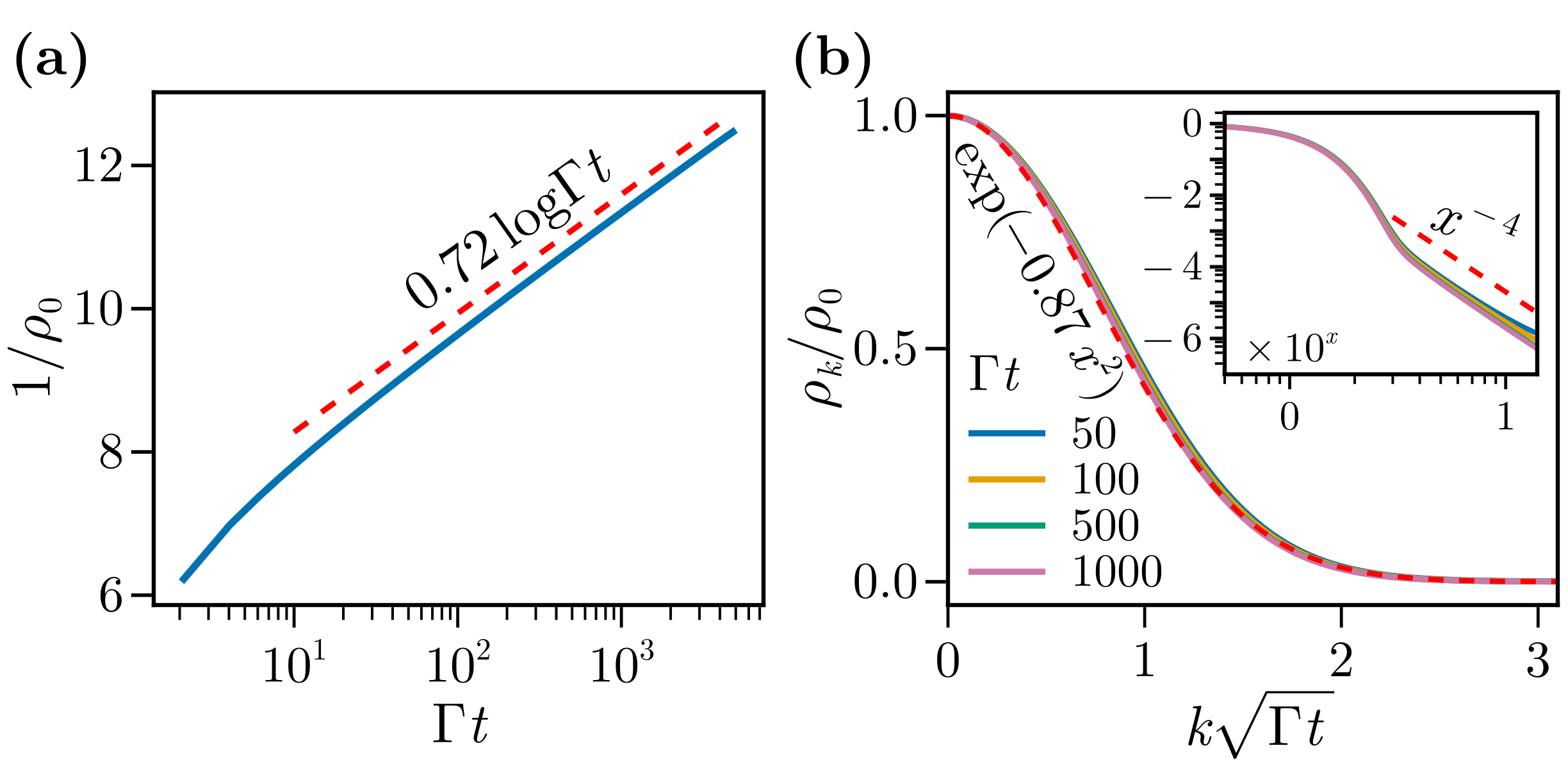}
    \caption{Long-time dynamics of fermionic densities starting from the fully excited state by numerically solving \cref{eq:dr/dt} exact in the limit $\Gamma/J\ll 1$. (a) The $k=0$ mode decays consistently with $\rho_0(t)^{-1}\sim (2/\pi)\log \Gamma t$. (b)~The normalized distribution $\rho_k(t)/\rho_0(t)$ shows a scaling collapse with $k\sqrt{\Gamma t}$ consistent with a scaling function $f(x)=e^{-x^2}$ for $|x|\lesssim 1$. (Inset) At larger momenta ($|x|\gtrsim 1$), this function decays as $1/x^4$. }
    \label{fig:rho_from_ODE}
\end{figure}

\textit{Analytic continuation.---}The kinetic equation is particularly complicated due to its \textit{nonlocal} structure in momentum space. Alternatively, the dynamics can be described elegantly by analytically continuing to the complex plane. We define an analytic function in the unit disk $|z|<1$ as
\begin{align}
    Q(z) = \int_0^{2\pi} \frac{{\rm d}k}{2\pi} \rho_k \frac{e^{ik}+z}{e^{ik}-z}\,.\label{eq:last_line}
\end{align}
One can then see that $\re Q(z= e^{ik - 0^+}) = \rho_k$ while $\im Q(e^{ik-0^+})={\cal H}[\rho_k]$ is the Hilbert transform of $\rho_k$. The analytic function $Q$ is time dependent, hence $Q=Q(t,z)$, and the resulting dynamics becomes a first-order partial differential equation in the complex plane \cite{sm}: 
\begin{align}
    &\partial_t Q +2m z\partial_z Q=\frac{1}{z}\Big\{-(1-z^2)n + (1-z)^2 Q \nonumber \\
    &-(1+z)^2 n^2+2(1-z^2) n Q  
    -  (1-z)^2  Q^2\Big\}  \,, \label{eq:Q_full} 
\end{align}
contrasted with the integro-differential equation involving $\rho_p$s. Here and in the rest of the paper, we absorb $\Gamma$ into time for ease of notation. 
This equation can be solved numerically by making, and truncating, a multipole expansion as $Q(t,z)=\sum_{p=0}^\infty a_p(t) z^p$  together with $n(t)=a_0(t)$ and $m(t)=2a_0(t)-a_1(t)$ and evolving the multipole coefficients in time.

\textit{Scaling regime.---}
Given its analytic form, \cref{eq:Q_full} enables further analytical progress. At long times, 
fermionic densities are concentrated around $k=0$, and are described by a scaling function of $k\sqrt{t}$ [see \cref{eq:scaling_rho_k}]. Thus, it is reasonable to identify a scaling region for $Q$ around $z=1$. With $z=1+w$ and $Q(t,z) \to Q(t,w)$, we can recast \cref{eq:Q_full} in terms of the scaling variable $w\sqrt{t}$, and drop 
terms that decay as additional powers of $1/t$. The resulting equation takes a rather simple form as 
\begin{equation}\label{eq:Q-dynamics}
    \partial_t Q +2m\partial_w Q= -(2n+wQ)(2n+wQ-w)\,.
\end{equation}
Interestingly, we would obtain this equation in the hard-core limit of the Lieb-Liniger gas with the dissipator taking the form $\hat L(x)\propto \partial_x \hat \Psi$ where $\hat\Psi$ is the bosonic annihilation operator. Just like our spin model, this model admits a dark state with a single bosonic excitation at $k=0$. However, the above equation is not completely well-defined without a UV cutoff (see also \cite{bouchoule_losses_2021}). In this sense, one could view \cref{eq:Q_full} as the UV completion of the dynamics in \cref{eq:Q-dynamics}. 
Nevertheless, \cref{eq:Q-dynamics} can be used as an intermediate step to organize our perturbative expansion at least to leading orders. 

As the dynamics of the zero mode is logarithmically slow [\cref{eq:rho0_log}], a reasonable ansatz for the scaling solution is given by 
\[
    Q(t,w) = \frac{1}{\log t}{\cal F}_1(w\sqrt{t}) + \frac{1}{(\log t)^2}{\cal F}_2(w\sqrt{t}) + \cdots,
\]
where ${\cal F}_i$s define scaling solutions as a function of $w\sqrt{t}$,
and can be solved order by order. It turns out that the scaling functions can be solved analytically, from which we also obtain analytical expressions for $\rho_k(t)= \re Q(w=ik-0^+) $. We leave the details to the SM \cite{sm}, and just report the asymptotic expressions:

\begin{equation}
    \rho_k(t) \sim 
    \begin{dcases}
        \frac{\pi}{2\log t} e^{-k^2 t}, & |k|\sqrt{t} \lesssim 1 \\
        \frac{\pi}{8(\log t)^2} \frac{1}{k^4 t^2}, & |k|\sqrt{t} \gg 1 \\
    \end{dcases}
\end{equation}
To the leading order in $1/\log t$, we thus find a Gaussian distribution in momentum identical to that of the free fermions, as promised. However, the universal logarithmic dependence (including the prefactor) is fixed by genuinely nonlinear terms in the dynamics \cite{sm}. 
Furthermore, the scaling $\rho_k \sim 1/k^4$  at large momenta is characteristic of hard-core interactions subject to lossy dynamics \cite{bouchoule_breakdown_2021}. What is particularly interesting here is that this term also takes a universal form, independent of initial conditions among other details.

So far, we have focused on fermionic densities $\rho_k$. 
Viewing spin-$\frac{1}{2}$ particles as hard-core bosons, we can also characterize the bosonic momentum distribution $n_k$. The latter can be computed from  a Toeplitz determinant constructed from fermionic correlators. We find that $n_k(t) \approx \rho_k(t)$ when $|k|\sqrt{t}\lesssim1$ \cite{sm}.  The fact that JW strings do not significantly contribute in this regime is because the average distance between spin excitations, $1/n(t) \sim \sqrt{t}\log t$, is larger than the characteristic length scale $\sqrt{t}$. The schematic picture in \cref{fig:schematic} thus holds for either fermionic or bosonic momentum distributions. 
We further find that $n_k$ also decays as $1/k^4$ at large momenta, and exhibits a scaling collapse with $k\sqrt{t}$ possibly with multiplicative logarithmic terms.

\begin{figure}[t!]
    \centering
    \includegraphics[width=1.0\linewidth]{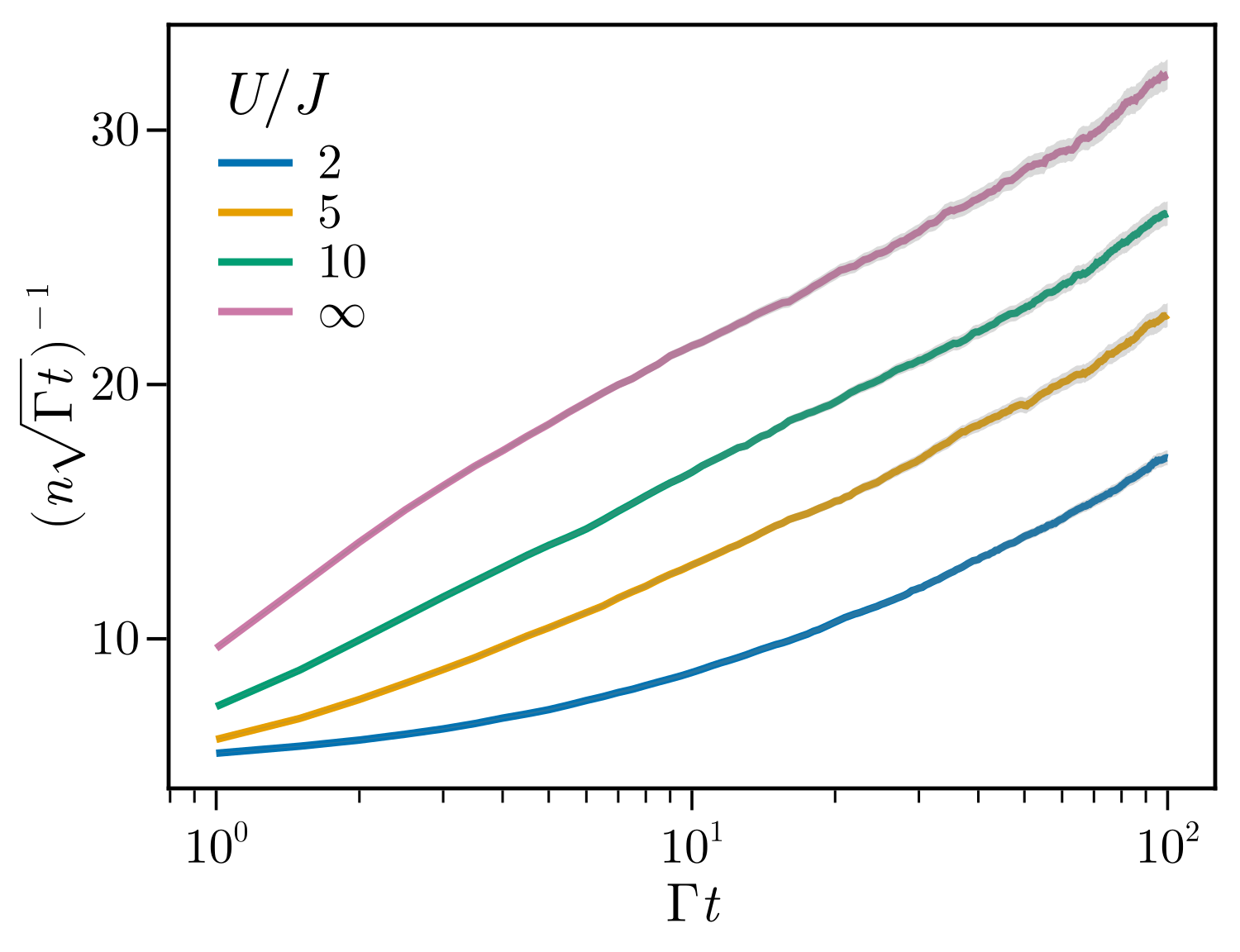}
    \caption{Universal dynamics for the dissipative Bose-Hubbard model. We plot the the rescaled inverse density $(n\sqrt{\Gamma t})^{-1}$ with on-site interactions $U/J = 2, 5, 10,\infty$; the latter corresponds to the XX spin chain (hard-core bosons). Data show scaling consistent with $\log \Gamma t$ at long times across different values of $U$. Simulations use $L=120$ sites with the initial state at half filling $n(0)=0.5$, loss rate $\Gamma/J = 0.5$, bond dimension $\chi = 16$, and time step $\delta t = 0.001$, averaged over $n_{\mathrm{traj}} \gtrsim 5000$ quantum trajectories. Error bands show  the standard error of the mean.}
    \label{fig:q_sim}
\end{figure}

\textit{Soft-core bosons.---}Next, we relax the hard-core constraint of the XX model, and consider a system of interacting bosons, $\hat H= -J \sum_j (\hat b_j^\dagger \hat b_{j+1}+{\rm h.c.}) + U\hat n_j(\hat n_j-1)/2$ together with the Lindblad operator $\hat L_j=\hat b_{j+1} - \hat b_j$ where $\hat b (\hat b^\dagger)$ denote bosonic operators, $\hat n_j=\hat b^\dagger_j \hat b_j$, and $U$ the strength of nonlinear interactions; the XX model is recovered in the limit $U\to\infty$ (with $J$ half of that in \cref{eq:Hamilt}). This model also admits a dark spin-wave excitation at $k=0$ controlling the slow dynamics. Interestingly, we find the same universal behavior at long times across different values of $U$. Using quantum trajectory MPS simulations, we show in \cref{fig:q_sim} that the total density decays as $n(t)\propto 1/\sqrt{\Gamma t} \log(\Gamma t)$. This universal behavior can be understood by recalling that, in 1D, the low-density limit of interacting bosons lies in the strongly interacting regime \cite{cazalilla_one_2011}, effectively recovering the hard-core constraint where the fermionic picture described here applies. In the opposite regime of large bosonic densities, quantum fluctuations are suppressed and a hydrodynamic approach becomes applicable, leading to KPZ-like dynamics \cite{baldwin_singularities_2021}. 

\textit{Conclusions and outlook.---}In this work, we have shown that a single-particle dark state qualitatively alters the many-body dynamics at long times, and identified its distinct universal nature. Specifically, the slow mode defined by this dark state is shown to decay slowly, due to nonlinear coupling to other modes, as $1/\log t$. An immediate direction for future research is to consider pumping, leading to a nontrivial steady state, or nonreciprocal coupling \cite{begg_quantum_2024,marche_open_2026}, and identify the resulting universal behavior. A particularly intriguing  avenue is to investigate situations where there are several, or even a continuum of, single-particle dark states, scenarios that arise  generically  in quantum optics platforms in various dimensions \cite{asenjogarcia_exponential_2017}. The emergent universal behavior in such setting could be far richer. Finally, it is desirable to identify types of dissipation and interactions that give rise to long-lived strongly correlated states of quantum matter.

\textit{Acknowledgments.---}We acknowledge useful discussions with  Darrick
Chang, Aash Clerk, and Andrew Pocklington. M.M. acknowledges support from the National Science Foundation under NSF CAREER Award No. DMR-2142866, and the Office of Naval Research grant N000142612232. R.D.~and J.S.~ are supported by the ERC Consolidator project MATHLOCCA (Grant nr.~101170485). They have received financing from the Interdisciplinary Thematic Institute QMat, as part of the ITI 2021-2028 program of the University of Strasbourg, CNRS and Inserm, and were supported by IdEx Unistra (ANR-10-IDEX-0002), and by SFRI STRAT’US project (ANR-20-SFRI-0012) and EUR QMAT ANR-17-EURE-0024 under the framework of the French Investments for the Future Program. They received further support from the French National Research Agency under the France 2030 program ANR-23-PETQ-0002 (PEPR project QUTISYM) and project ANR-21-ESRE-0032 (PEPR project aQCess). Computations  were  carried  out  using  resources  of  the High Performance Computing Center of the University of Strasbourg, funded by Equip@Meso (as part of the Investments for the Future Program) and CPER Alsacalcul/Big Data. A.S.N. was supported by
the Dutch Research Council (NWO/OCW) as a part
of the Quantum Software Consortium (project num-
ber 024.003.037), QDNL (project number
NGF.1582.22.030) and ENW-XL grant (project number
OCENW.XL21.XL21.122).


\makeatletter
\let\sm@save@addcontentsline\addcontentsline
\let\addcontentsline\@gobblethree   
\bibliography{refs.bib}
\let\addcontentsline\sm@save@addcontentsline
\makeatother

\ifincludesm
    \clearpage
    \onecolumngrid
    \ifincludesm
  \begin{center}
    \textbf{\large Supplemental Material:\\[0.3em] Universal dynamics from a single-particle dark state}
  \end{center}
  \vspace{1em}
\else
  \title{Supplemental Material for \\``Universal dynamics from a single-particle dark state''}
  \author{Ruben Daraban}
  \affiliation{CESQ/ISIS (UMR 7006), CNRS and Universit\'{e} de Strasbourg, 67000 Strasbourg, France}
  \author{Arghavan Safavi-Naini}
  \affiliation{Institute for Theoretical Physics, Institute of Physics, University of Amsterdam, Science Park 904, 1098 XH Amsterdam, The Netherlands}
  \affiliation{QuSoft, Science Park 123, 1098 XG Amsterdam, The Netherlands}
  \author{Johannes Schachenmayer}
  \affiliation{CESQ/ISIS (UMR 7006), CNRS and Universit\'{e} de Strasbourg, 67000 Strasbourg, France}
  \author{Mohammad Maghrebi}
  \affiliation{Department of Physics and Astronomy, Michigan State University, East Lansing, Michigan 48824 USA}
  \pacs{}
  \maketitle
\fi

\setcounter{secnumdepth}{3}   
\setcounter{figure}{0}
\setcounter{equation}{0}
\setcounter{section}{0}
\makeatletter
\renewcommand{\thefigure}{S.\@arabic\c@figure}
\renewcommand{\thesection}{S.\Roman{section}}
\renewcommand{\theequation}{S.\@arabic\c@equation}
\renewcommand{\thetable}{S.\@arabic\c@table}
\renewcommand{\theHfigure}{S\arabic{figure}}
\renewcommand{\theHequation}{S\arabic{equation}}
\renewcommand{\theHsection}{S\Roman{section}}
\renewcommand{\theHtable}{S\arabic{table}}
\makeatother

In this Supplemental Material,  we provide additional details on the results stated in the main text. 
In \cref{sec:fermions}, we provide the mapping to fermions for both the Hamiltonian and dissipative dynamics.  In \cref{sec:GGE}, we provide the details of the time-dependent generalized Gibbs ensemble constructed from fermions, derive the analytical equations reported in the main text, and discuss the dynamics in different limits. Finally, in \cref{sec:MPS}, we provide further details on the MPS simulations, provide additional numerical results and contrast against the analytical equations.

\tableofcontents

\section{Mapping to fermions}\label{sec:fermions}
In this section, we provide the mapping to fermions, and specifically derive \red{Eq.~(4)} of the main text. 
For convenience, we first apply a transformation $\sigma_i^{y,z}\to -\sigma_i^{y,z}$, a $\pi$ rotation around $x$. Under this transformation, the Hamiltonian remains unchanged, while the dissipator becomes
\begin{equation}\label{eq:newL}
    \hat L_j = \sigma^+_j - \sigma^+_{j+1}\,.
\end{equation}

Spins can be mapped to fermions via the Jordan-Wigner transformation: 
\begin{equation}
    \sigma^-_j=\prod_{l=1}^{j-1} (1-2\hat c_l^\dagger \hat c_l)\hat c^\dagger_j=e^{i\pi \sum_{l=1}^{j-1} \hat c_l^\dagger \hat c_l}\hat c_j^\dagger\,, \qquad \sigma^+_j= (\sigma^-_j)^\dagger\,, \qquad \sigma^z_j=1-2\hat c_j^\dagger \hat c_j\,.
\end{equation}
With this transformation, the Hamiltonian takes the form
\begin{align}
    H 
    = -2\sum_{j=1}^{L-1} J(\hat c^\dagger_j c_{j+1}+ c^\dagger_{j+1} c_{j})
     +2J e^{i\pi \hat N}(\hat c_L^\dagger c_1+ c_1^\dagger \hat c_L)\,,
\end{align}
where 
\begin{equation}
    \hat N = \sum_{j=1}^L \hat c_j^\dagger \hat c_j
\end{equation}
is the number operator.
Because the Hamiltonian is number-conserving, the even- and odd-parity sectors of $\hat H$ can be treated independently. 
One can write the Hamiltonian in the even and odd sectors as
\begin{align}\label{eq:two_sectors_Ham}
    \hat H_{\rm e/o} = -2J\sum_{j=1}^{L} \hat c^\dagger_j c_{j+1}+ {\rm h.c.}\,,
    \qquad \mbox{with} \quad \hat c_{L+1} =\mp \hat c_1\,. \nonumber
\end{align}
That is, the Hamiltonian becomes that of free fermions but with antiperiodic and periodic boundary conditions for the even and odd sectors, respectively. 
Going to Fourier space, we have
\begin{equation}\label{eq:Fourier-trans}
    c_{k}=\frac{1}{\sqrt{L}} \sum_{j=1}^L \hat c_j e^{-ikj}\,,
\end{equation}
with $k$  quantized as 
\begin{equation}
    k \in \mathbb Z_{\rm ap} =\frac{2\pi}{L}  (\mathbb Z_L+\frac{1}{2}), \qquad \mathbb Z_L = \{0,1, \cdots, L-1\} 
\end{equation}
in the even parity sector, while 
\begin{equation}
    k \in \mathbb Z_{\rm p}=  \frac{2\pi}{L}\mathbb Z_L 
\end{equation}
in the odd parity sector. 
The Hamiltonian in either sector is given by 
\begin{equation}
    H_{\rm e/o} = \sum_{k\in \mathbb Z_{\rm ap/p}} \omega_k c_{k}^\dagger c_{k}\,, \qquad \omega_k =- 2J \cos k\,.
\end{equation}

Next, we write the Lindblad operator in terms of fermions:
\begin{align}
    \hat L_j
    &=  \hat {\cal S}(j-1) \hat c_j -\hat {\cal S}(j) c_{j+1} = \hat {\cal S}(j) (e^{i\pi \hat c_j^\dagger \hat c_j}\hat c_j - \hat c_{j+1})\nonumber \\ 
    &=  \hat {\cal S}(j) (\hat c_j - \hat c_{j+1})\,,
\end{align}
where we have defined the string operator $\hat {\cal S}(j)= \exp(i\pi \sum_{l=1}^{j} \hat c_l^\dagger \hat c_l)$. 
Setting $j =L$ in the above equation, we have 
\begin{equation}
    \hat L_L=  \hat {\cal P} (\hat c_L - \hat c_{L+1})\,,
\end{equation}
with $\hat {\cal P} = \hat {\cal S}(L)$  the total parity. 
Expanding the rhs of this equation in  momentum space, we obtain
\begin{align}\label{eq:c12}
     \hat L_L=\frac{1}{\sqrt{L}} \hat {\cal P} \sum_k e^{ik L}(1- e^{ik}) \hat c_k = \frac{1}{\sqrt{L}} \sum_k(e^{ik}-1) \hat c_k \equiv \hat {\tilde c}\,,
\end{align}
where we have used $e^{ik L} = - \hat {\cal P}$ and $\hat {\cal P}^2=1$.  The above expression is particularly convenient because it allows us to write the Lindblad operator $\hat L_L$ linearly in fermionic operators.

Finally, we consider the full dynamics, ${\cal L}(\hat \rho) = -i[\hat H,\hat \rho] +{\cal D}(\hat \rho)$ where ${\cal D}(\cdot)$ describes the dissipative dynamics in terms of Lindblad operators, ${\cal D}(\cdot)= \Gamma \sum_ j \hat L_j^\dagger  \cdot \hat L_j -\frac{1}{2}(\hat L_j^\dagger \hat L_j \cdot +\cdot \hat L_j^\dagger \hat L_j )$.
While the Jordan-Wigner transformation maps the Hamiltonian to free fermions, the dissipator generally takes a complicated form. To be more precise, the term $\{\hat L_j^\dagger \hat L_j,\cdot\}$, which can be absorbed in a (non-hermitian) effective Hamiltonian, is mapped to a quadratic fermionic term; however, the jump term, $\hat L_j \cdot \hat L_j^\dagger$, produces fermionic string operators, rendering the resulting model highly nonlinear and nonlocal. However, one can still derive an exact equation for the dynamics of the total number of excitations, alternatively, the fermionic number operator, $\hat N= \sum_j (1-\sigma_j^z)/2=\sum_j \hat c_j^\dagger \hat c_j$. Note that 
\begin{equation}
    \frac{\rm d}{{\rm d}t}\langle \hat N\rangle =\tr(\hat N{\cal L}(\hat \rho))= \tr({\cal L}^{\ddagger} (\hat N)\hat \rho)=  \langle {\cal D}^\ddagger (\hat N) \rangle\,,
\end{equation}
where ${\cal L}^\ddagger$ is the adjoint Liouvillian,  ${\cal L}^\ddagger (\cdot) = i [\hat H, \cdot] +  {\cal D}^\ddagger(\cdot)$, with ${\cal D}^\ddagger(\cdot)= \Gamma\sum_ j \hat L_j^\dagger  \cdot \hat L_j -\frac{1}{2}(\hat L_j^\dagger \hat L_j \cdot +\cdot \hat L_j^\dagger \hat L_j )$. 
In the last equality in the above equation, we have used the fact that Hamiltonian is number conserving, $[\hat H,\hat N]=0$.
More conveniently, we can write ${\cal D}^\ddagger (\cdot) =\frac{\Gamma}{2}\sum_j \hat L_j^\dagger [\cdot, \hat L_j]+[\hat L_j^\dagger,\cdot]\hat L_j$. With $\hat L_j = \sigma^+_j - \sigma^+_{j+1}$, we have
\begin{equation}
    [\hat N, \hat L_j] = - \hat L_j\,, \qquad [ \hat N,\hat L_j^\dagger] =  \hat L_j^\dagger\,.
\end{equation}
Combining these equations, we find
\begin{equation}
    \frac{\rm d}{{\rm d}t}\langle \hat N\rangle  = -\Gamma \langle \sum_j \hat L_j^\dagger \hat L_j\rangle  = -\Gamma\langle \sum_j (\sigma^-_j - \sigma^-_{j+1})(\sigma^+_j - \sigma^+_{j+1})\rangle\,.
\end{equation}
Writing the spin operators in terms of fermions, we obtain
\begin{equation}
    \frac{\rm d}{{\rm d}t} \sum_j \langle \hat c_j^\dagger \hat c_j\rangle = -\Gamma \sum_j\langle (\hat c_{j+1}^\dagger -\hat c_{j}^\dagger)  (\hat c_{j+1} -\hat c_{j})\rangle\,.
\end{equation}
We thus recover \red{Eq.~(4)} of the main text.

\section{Perturbative approach at weak dissipation}\label{sec:GGE}
In this section, we provide a perturbative approach at weak dissipation based on the time-dependent generalized Gibbs ensemble (GGE) constructed from free fermions. In \cref{subsec:GGE}, we provide a general description of the perturbative approach. In \cref{subsec:GGE_fermions}, we specialize to our model, and formulate the GGE dynamics in terms of fermions. Specifically, we derive \red{Eqs.~(5,10,11)} of the main text. Furthermore, we compute the bosonic momentum distribution from its fermionic counterpart. Finally, in \cref{subsec:GGE_dynamics}, we investigate the solutions to the GGE equation in different limits and analytically derive and numerically verify \red{Eqs.~(7,8,12)} of the main text. 

\subsection{Time-dependent GGE (GE) at weak dissipation}\label{subsec:GGE}
To develop a perturbative approach at weak dissipation, let us first consider 
the dynamics of a (quasi-)local operator $\hat q_j$ with support possibly on several sites around site $j$. The expectation value of this operator changes over time as 
\begin{equation}
    \frac{\rm d}{{\rm d}t}\langle \hat q_j\rangle =\tr(\hat q_j{\cal L}(\hat \rho))\,.
\end{equation}
Assuming that the timescale associated with loss is long compared to the Hamiltonian dynamics, the density matrix can be written in the Hamiltonian eigenbasis as different energy levels dephase and no revivals occur in the thermodynamic limit $L\to \infty$. More precisely, for an integrable Hamiltonian, the density matrix can be described by a generalized Gibbs ensemble (GGE) ansatz \citeSM{lange_pumping_2017,lange_time_2018}, while for a generic Hamiltonian, it simply becomes a Gibbs ensemble (GE) \citeSM{Mori_thermalization_2020}.  With this ansatz, the Hamiltonian commutes with the density matrix and drops out of the dynamics.
At the same time, the temperature(s) describing the GE (GGE) slowly evolve in time due to dissipation.
Alternatively, the expectation value of the operator $\hat q_j$ evolves as 
\begin{equation}
   \frac{1}{\Gamma}\frac{\rm d}{{\rm d}t}\langle \hat q_j\rangle = \sum_{l} \tr \left[\hat q_j\left(\hat L_l\hat \rho \hat L_l^\dagger-\frac{1}{2}\{\hat L^\dagger_l \hat L_l,\rho\}\right)\right]\,.
\end{equation}
Assuming (lattice) translation invariance and defining $\hat Q=\sum_j \hat q_j$, the above equation becomes 
\begin{equation}
     \frac{1}{\Gamma}\frac{\rm d}{{\rm d}t}\langle \hat q_j\rangle =\tr \left[\hat Q\left(\hat L_1\hat \rho \hat L^\dagger_1-\frac{1}{2}\{\hat L^\dagger_1 \hat L_1,\hat\rho\}\right)\right]\,.
\end{equation}
Obviously, the choice of the first site is arbitrary, and one may instead choose any other site if convenient (as we will do later).
Finally, if $\hat Q$ is conserved then the density matrix commutes with this operator as well, and the above equation becomes \citeSM{bouchoule_effect_2020}
\begin{equation}\label{eq:star}
     \frac{1}{\Gamma}\frac{\rm d}{{\rm d}t}\langle \hat q_j\rangle =\tr \left[\left(\hat L^\dagger_1\hat Q \hat L_1-\hat Q \hat L^\dagger_1 \hat L_1\right)\hat \rho\right]\,.
\end{equation}

\subsection{GGE from free fermions}\label{subsec:GGE_fermions}

For purely Hamiltonian dynamics and for even observables (i.e., those not changing the parity), we can limit ourselves to the even sector \citeSM{calabrese_quantum_2012}.  
However, introducing dissipation violates parity conservation. 
Still the system will converge to a GGE state within each sector:
\begin{equation}
    \hat \rho= \hat P_+ \hat \rho+  \hat P_- \hat \rho= \hat \rho_{\rm GGE,+} + \hat \rho_{\rm GGE,-}\,,
\end{equation}
where $\hat P_\pm =\left(1\pm (-1)^{\hat N}\right)/2$ is the projector on the even/odd sector. Note that $\hat \rho_{\rm GGE,\pm}$ are Gaussian states which allow us to use the Wick's theorem. 
Defining the mixed projector $\hat P_k \equiv  \hat P_+ \delta_{k\in \mathbb Z_{\rm ap}} + \hat P_- \delta_{k\in \mathbb Z_{\rm p}}$, the conserved charges are given by 
\begin{equation}\label{eq:Qbb}
    \hat Q_k = L \hat P_k \hat c_k^\dagger \hat c_k\,, \qquad \rho_k = \langle \hat P_k \hat c_k^\dagger \hat c_k\rangle\,.
\end{equation}
Notice that $\rho_k$ is the fermionic density, not the density matrix.

\subsubsection{Basis transformation}
In the following analysis, we also need a change of basis states between the even and odd sectors. 
Since both $\hat c_k$ for $k \in \mathbb Z_{\rm ap}$ and $\hat c_p$ for $p \in \mathbb Z_{\rm p}$ span the one-particle Hilbert space on a ring, they can be related to each other. Indeed, the corresponding momentum eigenstates are related via 
\begin{equation}
    |k\rangle =\sum_{p\in \mathbb Z_{\rm p}} \langle p |k\rangle |p\rangle\,,
\end{equation}
where 
\begin{equation}
    \langle p |k\rangle =\frac{1}{L}\sum_{j=1}^L e^{i(k-p)j}=-\frac{2}{L}\frac{1}{1-e^{-i(k-p)}}\,.
\end{equation}
Identifying $|k\rangle=\hat c^\dagger_{k}|0\rangle$, we find 
\begin{equation}
    \hat c_k = -\frac{2}{L}\sum_{p\in \mathbb Z_{\rm p}}\frac{1}{1-e^{i(k-p)}} \hat c_p\,.
    \label{eq:basis_tran}
\end{equation}
Up to a relative sign, this expression is given in Ref.~\citeSM{Iorgov2011}. 
Taking the thermodynamic limit $L\to \infty$, we recover the basis transformation 
\(
    \hat c_k = -\frac{2i}{L}\sum_{p_m\in \mathbb Z_{\rm p}}\frac{1}{k_n-p_m} \hat c_p
\)
reported in \citeSM{bouchoule_effect_2020}. Finally, from the completeness relation
\begin{equation}
    \sum_{p\in \mathbb Z_{\rm p}} \langle k'| p \rangle \langle p|k\rangle =\delta_{k,k'}\,,
\end{equation}
we arrive at the condition 
\begin{equation}\label{identity}
    \sum_{p\in \mathbb Z_{\rm p}} \frac{1}{1-e^{-i(k-p)}} \frac{1}{1-e^{i(k'-p)}} =\frac{L^2}{4}\delta_{k, k'}\,.
\end{equation}

\subsubsection{Wick's theorem}

Here, we use \cref{eq:star} but substitute $\hat L_1$ with $\hat L_L$ for convenience to find
\begin{align}\label{eq:star2_corr}
    \frac{1}{\Gamma} \frac{\rm d}{{\rm d}t}\rho_k =\langle \hat {\tilde c}^\dagger  \hat Q_k  \hat {\tilde c}\rangle-\langle  \hat Q_k  \hat {\tilde c}^\dagger\hat {\tilde c} \rangle\,.
\end{align}
where the operator $\hat {\tilde c}$, defined in \cref{eq:c12}, is linear in fermionic operators. 
The second term on the rhs of the above equation does not involve a change of parity ($\hat {\tilde c}^\dagger \hat {\tilde c}$ conserves the
particle number), and can be easily computed using the Wick’s theorem as 
\begin{align}\label{eq:1st_term}
    \langle \hat P_k  \hat {\tilde c}^\dagger \hat {\tilde c} \hat c_k^\dagger \hat c_k\rangle =m \rho_k+ \frac{1}{L}\left|e^{ik}-1\right|^2\rho_k(1-\rho_k)\,,
\end{align}
where we have used  $\{\hat {\tilde c},c^\dagger_k\}=(e^{ik}-1)/\sqrt{L}$, and 
\begin{equation}\label{eq:n_2}
    \langle \hat {\tilde c}^\dagger \hat {\tilde c}\rangle=\int \frac{{\rm d}k}{2\pi} \left|e^{ik}-1\right|^2 \rho_k = \frac{1}{\pi}\int {\rm d}k (1-\cos k) \rho_k=m\,.
\end{equation}

We then consider the first term on the rhs of \cref{eq:star2_corr}. This term involves a change of the sector due to the jump term, so we first make a basis transformation 
\begin{equation}
    \langle \hat {\tilde c}^\dagger \hat P_k \hat c_k^\dagger \hat c_k \hat {\tilde c} \rangle =\frac{4}{L^2}\sum_{\lambda,\lambda'} \frac{1}{1-e^{-i(k-\lambda)}}\frac{1}{1-e^{i(k-\lambda')}} \langle \hat {\tilde c}^\dagger c^\dagger_\lambda c_{\lambda'} \hat {\tilde c}\rangle \,.
\end{equation}
Next, we can apply the Wick's theorem:
\begin{align}
    \langle \hat {\tilde c}^\dagger c^\dagger_{\lambda} c_{\lambda'} \hat {\tilde c}\rangle  
    =
   m \delta_{\lambda\lambda'}\rho_\lambda-\frac{e^{-i\lambda'}-1}{\sqrt{L}}\rho_{\lambda'}\frac{e^{i\lambda}-1}{\sqrt{L}} \rho_\lambda
\end{align}
Combining the last two equations, we find
\begin{align}
    \langle \hat {\tilde c}^\dagger \hat P_k \hat c_k^\dagger \hat c_k \hat {\tilde c} \rangle &=\frac{4}{L^2}\left(m\sum_\lambda {\frac{\rho_\lambda}{|1-e^{i(k-\lambda)}|^2}}-\frac{1}{L}\left|\sum_\lambda\frac{(e^{i\lambda}-1) \rho_\lambda}{1-e^{-i(k-\lambda)}}\right|^2\right) \nonumber \\
    &=\frac{4}{L^2}\left(m\sum_\lambda {\frac{ \rho_\lambda}{|e^{ik}-e^{i\lambda}|^2}}-\frac{1}{L}\left|\sum_\lambda\frac{(e^{i\lambda}-1) \rho_\lambda}{e^{ik}-e^{i\lambda}}\right|^2\right)\label{eq:line_2}
\end{align}
To remove the divergence in the first term on the second line, we write 
\begin{equation}
    \sum_\lambda \frac{\rho_\lambda}{|e^{ik}-e^{i\lambda}|^2}=\sum_\lambda \frac{\rho_\lambda-\rho_k}{|e^{ik}-e^{i\lambda}|^2}+\sum_\lambda\frac{1}{|e^{ik}-e^{i\lambda}|^2}\rho_k= \sum_\lambda \frac{\rho_\lambda-\rho_k}{|e^{ik}-e^{i\lambda}|^2}+\frac{L^2}{4} \rho_k\,,
\end{equation}
using the completeness relation, \cref{identity}. 
Then, \cref{eq:line_2} can be written as 
\begin{align}\label{eq:2nd_term}
    \langle \hat {\tilde c}^\dagger \hat P_k \hat c_k^\dagger \hat c_k \hat {\tilde c} \rangle =\frac{4}{L} \left(m \dint \frac{{\rm d}\lambda}{2\pi} \frac{\rho_\lambda-\rho_k}{|e^{ik}-e^{i\lambda}|^2}-\left|\dint \frac{{\rm d}\lambda}{2\pi}\frac{(e^{i\lambda}-1)\rho_\lambda}{e^{ik}-e^{i\lambda}}\right|^2\right)+ m\rho_k
\end{align}
where we have taken the thermodynamic limit, substituting $\sum_\lambda \cdot= \frac{L}{2\pi}\int_{-\pi}^{\pi}{\rm d}k \, \cdot$; the slash denotes the Cauchy principal value of the integral. Combining \cref{eq:1st_term,eq:2nd_term}, we finally obtain
\begin{subequations}\label{eq:final-4}
\begin{align}
    &\frac{1}{\Gamma}\frac{\rm d}{{\rm d}t}\rho_k= -F_k\,,
\\
&\mbox{with} \qquad F_k= \left|e^{ik}-1\right|^2\rho_k - \left(\left|e^{ik}-1\right|^2\rho_k^2-\left|\frac{1}{\pi}\dint d\lambda \frac{(e^{i\lambda}-1)\rho_\lambda}{e^{ik}-e^{i\lambda}}\right|^2\right) 
+\frac{2m}{\pi} \dint d\lambda \frac{\rho_k-\rho_\lambda}{|e^{ik}-e^{i\lambda}|^2}\,. \label{eq:F_rho_k_sm}
\end{align}
\end{subequations}
This concludes our derivation of \red{Eq.~(5)} of the main text. 

The terms in \cref{eq:F_rho_k_sm} can be reorganized into a form that admits an analytic continuation to the complex plane. We start from the term involving the double integral, rewriting it as
\begin{align*}
    &\left| \dint \frac{{\rm d}\lambda}{\pi} \frac{(1-e^{i\lambda})\rho_\lambda}{1-e^{i(\lambda-k)}}\right|^2 = \left| \dint \frac{{\rm d}\lambda}{2\pi} \big((1-\cos \lambda)-i\sin\lambda\big)\big(1+i \cot\frac{\lambda-k}{2}\big)\rho_\lambda\right|^2  \\
    =& \left( \dint \frac{{\rm d}\lambda}{2\pi} \rho_\lambda (1-\cos \lambda + \sin\lambda \cot\frac{\lambda-k}{2})  \right)^2 + \left( \dint \frac{{\rm d}\lambda}{2\pi} \rho_\lambda  (-\sin\lambda + (1-\cos \lambda)\cot\frac{\lambda-k}{2})\right)^2   \\
    =& \left(\frac{m}{2}+ \dint \frac{{\rm d}\lambda}{2\pi} \rho_\lambda  \sin\lambda \cot\frac{\lambda-k}{2} \right)^2 + \left( \dint \frac{{\rm d}\lambda}{2\pi} \rho_\lambda  (1-\cos\lambda) \cot\frac{\lambda-k}{2}\right)^2\,,
\end{align*}
where we have used $\frac{1}{\pi}\int {\rm d}k \rho_k(1-\cos k)=m$ [see \cref{eq:n_2}] and $\int {\rm d}k \rho_k\sin k=0$ due to ($k\to -k$) symmetry, and also used the identity
\begin{equation}
    \frac{1}{1-e^{i\lambda}}=\frac{1}{2}\big(1+ i \cot\frac{\lambda}{2}\big)\,.
\end{equation}
Finally, the last term in \cref{eq:F_rho_k_sm} is the Hadamard finite part and can be written as
\begin{equation}\label{eq:Hadamard_finite_part}
     \dint \frac{{\rm d}\lambda}{2\pi} \frac{\rho_k-\rho_\lambda}{1-\cos(\lambda-k)} = -\partial_k \dint \frac{{\rm d}\lambda}{2\pi} \rho_\lambda \cot\big( \frac{\lambda-k}{2}\big)\,,
\end{equation}
where we have used 
\begin{equation}
    \partial_k \cot\big( \frac{\lambda-k}{2}\big)=\frac{1}{1-\cos(\lambda -k)}\,.
\end{equation}
We can then rewrite \cref{eq:F_rho_k_sm} as 
\begin{align}\label{eq:Frho_4}
\begin{split}
F_k=&2(1-\cos k) \rho_k +\frac{m^2}{4} +m \dint \frac{{\rm d}\lambda}{2\pi} \sin \lambda\rho_\lambda\cot\big(\frac{\lambda-k}{2}\big) \\
&- \left(\big((1-\cos k)\rho_k\big)^2-\left(\dint \frac{{\rm d}\lambda}{2\pi} (1-\cos \lambda)\rho_\lambda\cot\big(\frac{\lambda-k}{2}\big)\right)^2\right) \\
& 
- \left(\big(\sin k\rho_k\big)^2-\left(\dint \frac{{\rm d}\lambda}{2\pi} \sin \lambda\rho_\lambda\cot\big(\frac{\lambda-k}{2}\big)\right)^2\right) \\
&-2m \partial_k \dint \frac{{\rm d}\lambda}{2\pi} \rho_\lambda \cot\big( \frac{\lambda-k}{2}\big)\,.
\end{split}
\end{align}
This expression is particularly convenient for the analytic continuation as the circular Hilbert transform of a function $A_k$, defined as ${\cal H}[A_k]= \dint \frac{{\rm d}\lambda}{2\pi} A_\lambda \cot(\frac{k-\lambda}{2})$, appears in all the integrals. 

\subsubsection{Analytic continuation}
As defined in \red{Eq. (9)} of the main text, one can make an analytic continuation to the unit disk via 
\begin{align}\label{eq:circ_HT}
    {Q}_A(z) = \int_0^{2\pi} \frac{{\rm d}k}{2\pi} A_k \frac{e^{ik}+z}{e^{ik}-z} \quad \Rightarrow \quad {Q}_A(z) \mbox{\, is analytic in \, } |z|<1\,, \quad \mbox{and}\quad  \re {Q}_A(e^{ik - 0^+}) = A_k\,.
\end{align}
A Kramers-Kronig relation follows as well:
\begin{equation}
    \im {Q}_A(e^{ik - 0^+}) = \dint \frac{{\rm d}\lambda}{2\pi} A_\lambda \cot(\frac{k-\lambda}{2})\,,
\end{equation}
thus reducing to the circular Hilbert transform of the function $A_k$. 
Furthermore, it is easy to see that 
\begin{equation}
    \re ({Q}_A(e^{ik - 0^+})^2) = A_k^2 - \left(\dint \frac{{\rm d}\lambda}{2\pi} A_\lambda \cot(\frac{k-\lambda}{2})\right)^2\,,
\end{equation}
and
\begin{equation}
   \re z\partial_z Q_A(z)\Big|_{z=e^{ik-0^+}} =  \partial_k \left( \im Q_A(z)\Big|_{z=e^{ik-0^+}}\right)= - \partial_k \dint \frac{{\rm d}\lambda}{2\pi} A_\lambda \cot\big( \frac{\lambda-k}{2}\big)\,.
\end{equation}

An analytic continuation of \cref{eq:Frho_4} then yields 
\begin{align}\label{eq:Qs}
    \partial_t {Q} = -[2{Q}_c + \frac{m^2}{4} + im {Q}_s- {Q}_c^2- {Q}_s^2+2m z\partial_z {Q}]\,,
\end{align}
where we have defined ${Q}, {Q}_c, {Q}_s$ as the analytic continuation of $\rho_k, (1-\cos k)\rho_k, \sin k \rho_k$, respectively. It is straightforward to write the functions ${Q}_c,{Q}_s$ in terms of ${Q}$. To do so, it is convenient to define 
\begin{equation}\label{eq:Qpm}
    {Q}_\pm(z)\equiv \int \frac{{\rm d}k}{2\pi} e^{\pm ik}\rho_k \frac{e^{ik}+z}{e^{ik}-z} = Q(z)-{Q}_c(z) \pm i {Q}_s(z)\,.
\end{equation}
Some manipulation gives
\begin{align}\label{eq:Q+}
    {Q}_+(z) =& \int \frac{{\rm d}k}{2\pi} \rho_k [(e^{ik}-z)+z]\frac{e^{ik}+z}{e^{ik}-z} = \int  \frac{{\rm d}k}{2\pi} \rho_k (e^{ik}+z) + z {Q}(z) \nonumber \\
    = &\,\, n -\frac{m}{2} + nz + z {Q}(z) \,,
\end{align}
where we have used 
\begin{equation}
    \int \frac{{\rm d}k}{2\pi} \rho_k =n, \qquad  \int \frac{{\rm d}k}{2\pi} e^{\pm ik } \rho_k = n -\frac{m}{2} \,.
\end{equation}
Similarly, we have 
\begin{align}\label{eq:Q-}
    {Q}_- (z)=& \int \frac{{\rm d}k}{2\pi} \rho_k [(e^{-ik}z -1)/z + 1/z ]\frac{1+e^{-ik}z}{1-e^{-ik}z} = -\frac{1}{z}\int \frac{{\rm d}k}{2\pi} \rho_k (1+e^{-ik}z) +\frac{1}{z}{Q}(z) \nonumber \\
    =&  - n+\frac{m}{2} + \frac{{Q}(z) -n}{z} \,.
\end{align}
While it might appear that the above equation has a simple pole at $z=0$, we note that 
\begin{equation}
    \lim_{z\to 0} {Q}(z)=n\,,
\end{equation}
hence, the residue is zero. Solving $Q_{c,s}$ in terms of $Q$ using \cref{eq:Qpm,eq:Q+,eq:Q-},
\cref{eq:Qs} can be written purely in terms of $Q$, and we recover \red{Eq.~(10)} of the main text. 
Notice, perhaps surprisingly, the function $m(t)$ has disappeared except in the last term on the rhs of the latter equation. 

The function $Q(z)$ can be expanded as a multipole series,  $Q(z) = \sum_{p=0}^\infty a_p z^p$. The fermionic density is then given by $\rho_k=\sum a_p\cos(k p)$, therefore, $a_0= \int \frac{{\rm d}k}{2\pi}\rho_k = n$ and $a_p = \int \frac{{\rm d}k}{\pi} \rho_k \cos(kp)$ for $p\ge1$.  
Specifically, for $p=1$, we have $a_1 = \int \frac{{\rm d}k}{\pi} \cos k =2n +  \int \frac{{\rm d}k}{2\pi} \rho_k |e^{ik}-1|^2 = 2n + m$. 
Finally, we remark that, expanding both sides of \red{Eq.~(10)} of the main text to the zeroth order in $z$, one finds that $\frac{1}{\Gamma}\dot a_0= 2n-a_1$. Substituting for $a_0, a_1$ from these expressions, we find $\frac{1}{\Gamma}\dot n = -m$, thus we recover the exact relation in \red{Eq.~(4)} of the main text, though in the weak dissipation limit. 

\subsubsection{Dissipative Tonks-Girardeau gas}\label{sec:continuum}
Here, we briefly consider the dissipative Tonks-Girardeau gas. We start from the Lieb-Liniger Hamiltonian
\begin{equation}
    \hat H = \int_x \hat \Psi^\dagger (-\partial_x^2/2 + g \hat \Psi^\dagger \hat \Psi) \hat \Psi\,,
\end{equation}
where $\hat \Psi, \hat \Psi^\dagger$ denote the bosonic field operators satisfying the commutation relation $[\hat \Psi(x),\hat \Psi^\dagger(y)]=\delta(x-y)$;  we have set $\hbar = m=1$.
The hard-core limit $g\to \infty$ recovers the Tonks-Girardeau gas. 
The resulting Hamiltonian then becomes the continuum limit of the lattice XX model. In the same limit, the Lindblad operator is given by 
\begin{equation}
    \hat L(x)= \partial_x \hat \Psi\,.
\end{equation}
In momentum space, we have $\hat L_k \sim k \hat \Psi_k$, so the dissipation rate increases with $k$ without bound. The resulting dynamics is thus not fully well-defined, and should be regularized. The lattice model exactly provides such regularization. However, it is useful to consider the continuum model where long wavelength features of the dynamics can be extracted. 
In the absence of dissipation, much like the XX model, hard-core bosons in 1D map to free fermions, and can be solved exactly \citeSM{cazalilla_one_2011}. In the presence of dissipation, this is no longer the case, and the quantum jump terms again pick up infinite Jordan-Wigner tails upon mapping to fermions. Nevertheless, we can utilize the same perturbative approach assuming weak dissipation. The resulting analysis closely follows the steps for the XX model with the simple prescription of expanding exponential functions such as $e^{ik}$ to first order in $k$. This is because in going to the continuum the wavevector $|k|\ll 1$ is much smaller that the lattice cutoff. The end result is thus given by \cref{eq:F_rho_k_sm} once all the exponential factors are expanded to the first order. One can then proceed with an analytic continuation to the upper half plane as \citeSM{bouchoule_effect_2020}
\begin{equation}
    Q(w)\equiv \frac{i}{\pi}\int {\rm d}k  \frac{\rho_k }{w-k}, \qquad \im w>0
\end{equation} 
whose dynamics then takes the form of a purely local analytic equation in $w$.
The resulting analytic equation coincides exactly with \red{Eq.~(12)} of the main text upon transforming $w\to -iw $ in the latter equation (to map the region defined by $\re w <0$ to the upper half plane, $\im w>0$). 
As pointed out earlier, this model is not completely well-defined; for example, a multipole expansion in $1/w$ breaks down beyond leading order. However, it is still useful to extract universal properties at long wavelengths in the vicinity of $w=0$. 

\subsection{Dynamics from the GGE equation}\label{subsec:GGE_dynamics}
Having established an analytical equation that governs the dynamics in the weak dissipation limit, here we provide analytical and numerical solutions in several limits. 

\subsubsection{Analytical solution at long times}
In this and the next subsection, we set $\Gamma=1$ for convenience. We start with \red{Eq.~(6)} of the main text at long wavelengths ($1/|e^{ip}-1|^2\approx p^2$): 
\begin{equation}\label{eq:6longwavelenth}
    \frac{\rm d}{{\rm d}t}\rho_0(t)=-\frac{1}{\pi^2}\left|\int {\rm d}\lambda \rho_\lambda(t) \right|^2-\frac{2m}{\pi}\int \frac{\rho_0(t)-\rho_\lambda(t)}{\lambda^2}\,,
\end{equation}
together with the scaling solution $\rho_k(t) = \rho_0(t) f(k\sqrt{t})$ [\red{(Eq.~(7)} of the main text)] to obtain an analytical solution for $\rho_0(t)$. We first note that 
\begin{equation}
    n(t)=\int \frac{{\rm d}k}{2\pi} \rho_k(t) = \frac{I}{2\pi} \frac{\rho_0(t)}{\sqrt{t}}\,, \qquad m(t)=-\dot n =\frac{I}{2\pi} \frac{\rm d}{{\rm d}t}\left(\frac{\rho_0(t)}{\sqrt{t}}\right)\,,
\end{equation}
where $I\equiv\int_{-\infty}^\infty {\rm d}xf(x)$. 
Substituting the scaling solution in \cref{eq:6longwavelenth}, we find
\begin{equation}
    \frac{\rm d}{{\rm d}t}\rho_0(t)= \frac{1}{\pi^2}\left[-\frac{I^2}{t}\rho_0(t)^2+I \tilde I \frac{\rm d}{{\rm d}t}\left(\frac{\rho_0(t)}{\sqrt{t}}\right)\rho_0(t)\sqrt{t}\right]\,,
\end{equation}
where $\tilde I= \int_{-\infty}^\infty {\rm d}x \frac{1-f(x)}{x^2}$. This equation can be then solved for $\rho_0(t)\equiv a(t)$ as
\begin{equation}
    \frac{1}{a(t)}+B \log \frac{a(t)}{a(\tau)}=\frac{1}{a(\tau)}+ (A+\frac{B}{2})\log\frac{t}{\tau}\,,
\end{equation}
where $\tau$ denotes the onset of the scaling solution, and  $A=I^2/\pi^2$ and $B=I\tilde I/\pi^2$. 
The exact solution of this equation can be written as the inverse of a ProductLog function (in the conventions of Mathematica).
If $a(t),a(\tau)$ are sufficiently small, we can ignore the second term on the lhs of the above equation to obtain 
\begin{equation}\label{eq:logt/tau}
    a(t)\sim  \frac{1}{a(\tau)^{-1}+(A+B/2)\log t/\tau}\,.
\end{equation}
For $f(x)\sim e^{-x^2}$, which will be justified in the next subsection, the coefficient of the logarithm in the denominator becomes $2/\pi$. 

\subsubsection{Full scaling solution}
To identify the full scaling solution, it is more convenient to work with the analytically continued function $Q(t,z)$ in \red{Eq.~(10)} of the main text and expand it around $z=1$. Let $z=1+w$ and, in an abuse of notation, $Q(t,z)\to Q(t,w)$ in the new coordinates. To the first nontrivial order derived in the previous section, we have $\rho_k(t)\propto  \frac{1}{\log t} f(k\sqrt{t})$, hence, to the same order, $Q$ scales as $1/{\log t}$, up to a scaling function in $w\sqrt{t}$. Recasting \red{Eq.~(5)} in terms of this scaled variable, and dropping terms that decay at higher orders in $1/t$,
we find
\begin{equation}\label{eq:Q-dynamics_sm}
    \partial_t Q = 2wn+w^2Q-(2n+wQ)^2-2m(t)\partial_w Q\,.
\end{equation}
One obtains exactly the same equation from the dissipative Tonks-Girardeau model discussed in \cref{sec:continuum}. Again, this equation is not completely well-defined; for example, a multipole expansion in powers of $1/w$ breaks down in the subleading order. However, expanding in the vicinity of $w=0$, this equation appears to be well-defined in the sense described below. 

\Cref{eq:Q-dynamics_sm} still cannot be solved exactly; however, an analytical solution can be constructed perturbatively, with time controlling the order of perturbation theory. Our earlier analysis motivates a series expansion as
\begin{equation}\label{eq:Q_ansatz}
    Q(t,z)\to  Q(t,w) = \frac{1}{\log t}{\cal F}_1(w\sqrt{t}) + \frac{1}{(\log t)^2}{\cal F}_2(w\sqrt{t}) + \cdots\,.
\end{equation}
This dictates a similar structure for the fermionic densities,
\begin{equation}
    \rho_k(t) = \frac{1}{\log t}{f}_1(k\sqrt{t}) + \frac{1}{(\log t)^2}{f}_2(k\sqrt{t}) + \cdots\,,
\end{equation}
where we have defined $f_i(x)=\re {\cal F}_i(0^- + i x)$. We can similarly expand the total density
\begin{equation}\label{eq:n_ansatz}
    n(t) \sim \left(\frac{A_1}{\log t}+\frac{A_2}{(\log t)^2}+\cdots\right)\frac{1}{\sqrt{t}}, \qquad m= -\dot n\,,
\end{equation}
with
\begin{equation}\label{eq:A_from_F}
    A_i = \int \frac{{\rm d}x}{2\pi} \re {\cal F}_i(0^- + i x) = \int \frac{{\rm d}x}{2\pi} {f}_i( x) \,.
\end{equation}
Here, we can safely extend the limits of the integral to $(-\infty,\infty)$ assuming that the  density decays sufficiently fast as $|x|\to\infty$. 

Inserting the ansatz in \cref{eq:Q_ansatz,eq:n_ansatz} in \cref{eq:Q-dynamics_sm}, we can solve for ${\cal F}_i$ order by order. To the first order, we find 
\begin{align}
\begin{split}
    &\frac{1}{2}{\cal F}_1'(u) - u {\cal F}_1(u)= 2A_1\quad \\ \Rightarrow \quad &{\cal F}_1 = 2\sqrt{\pi}A_1 e^{u^2}\left(1+{\rm Erf}(u)\right)\,, \quad f_1(x) = \re {\cal F}_1(0^- + i x) = 2\sqrt{\pi}A_1 e^{-x^2}\,,
\end{split}
\end{align}
where ${\rm Erf}(u)$ is the error function.
In solving this first-order differential equation, we have fixed the constant of integral by using \cref{eq:A_from_F}. This solution confirms our earlier statement that, to the first order, the scaling function is given by $f(x)\sim e^{-x^2}$. Notice however that to this order the constant $A_1$ remains undetermined. Our analytical solution in \cref{eq:logt/tau} in principle fixes this constant. But, we determine this constant   systematically within our perturbation theory, as we discuss next. 

Going to the next order, we find 
\begin{equation}
\frac{1}{2}\,u\,\mathcal{F}_2'(u)-u^2\,\mathcal{F}_2(u)
\;=\;
{-4A_1^2 + 2A_2 u}
+(1-4A_1 u)\,\mathcal{F}_1(u)
- u^2\,\mathcal{F}_1(u)^2
-A_1\,\mathcal{F}_1'(u)\,.
\end{equation}
Using the explicit solution for ${\cal F}_1(u)$, we can solve for ${\cal F}_2(u)$ as
\begin{equation}
    {\cal F}_2(u) = C_2 e^{u^2} + e^{u^2} \int_0^u dv  {\cal G}(v)\,,
\end{equation}
where $G(v)$ takes a rather complicated form, 
\begin{equation}
{\cal G}(v) =\frac{\pi(1+{\rm Erf}(v))}{v}
-\frac{3\pi\sqrt{\pi}}{2}\bigl(1+{\rm Erf}(v)\bigr)
-\frac{\pi e^{-v^{2}}}{v}
+4A_2e^{-v^{2}}
-\frac{\pi^2 v}{2}e^{v^{2}}\bigl(1+{\rm Erf}(v)\bigr)^{2}\,,
\end{equation}
but, more importantly, appears to diverge as $v\to 0$,
\begin{equation}\label{eq:v->0}
    {\cal G}(v) \sim (-8A_1^2 +2\sqrt{\pi}A_1 ) \frac{1}{v} + \cdots\,, \qquad v\to 0\,.
\end{equation}
This in turn gives rise to a logarithmic singularity of $\rho_k(t)$ which is nevertheless unphysical since $0\le\rho_k(t)\le 1$. We thus require the coefficient of the singular term in \cref{eq:v->0} to vanish, which then yields 
\begin{equation}
    A_1 \to \frac{\sqrt{\pi}}{4}\quad \Rightarrow \quad f_1(x)=\frac{\pi}{2}e^{-x^2}\,.
\end{equation}
We thus obtain the universal coefficient as expected. 

We can then determine the corresponding momentum distribution at this order: 
\begin{equation}\label{eq:C2+g}
    f_2(x) = \re {\cal F}_2(0^-+ix) = C_2 e^{-x^2} + g(x)\,,
\end{equation}
where $g(x)$ is given by 
\begin{equation}
    g(x)= e^{-x^2}\re \tilde g(i x)\,, \qquad \tilde g(u)= -\frac{\pi}{2} \int_0^{u} dv \frac{\pi  e^{v^2} y^2 \left(\text{Erf}(v)^2+1\right)+3 \sqrt{\pi } v \text{Erf}(v)-2 (1-e^{-v^2})}{v}\,.
\end{equation}
Notice that the constant $A_2$ drops out while $C_2$ still appears in \cref{eq:C2+g}. 
\begin{figure}[t!]
    \begin{center}
    \includegraphics[width=0.3\linewidth]{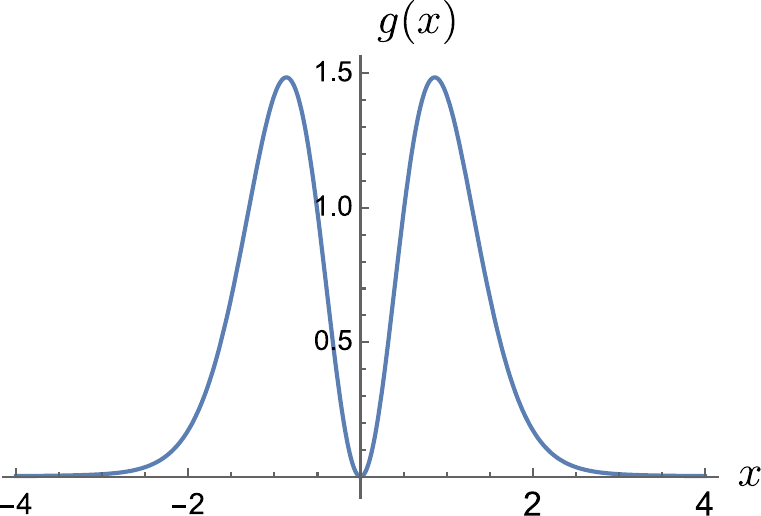}
    \caption{The function $g(x)$ defined in \cref{eq:C2+g}.}
    \label{fig:g(x)}
    \end{center}
\end{figure}
The above integral can be written in closed form, although it may not be illuminating. We plot this function in \cref{fig:g(x)}. Comparing to the Gaussian function in the first term on the rhs of \cref{eq:C2+g}, the function $g(x)$ is zero at $x=0$, peaks around $|x|\sim 1$, and, in contrast with the quick decay of the Gaussian function, falls off algebraically at large $|x|$ as 
\begin{equation}
    g(x)\sim \frac{\pi}{8x^4}\,, \qquad |x| \gg 1\,.
\end{equation}
See also \red{Eq.~(12)} of the main text.
While the constant $C_2$ in the above equation is still undetermined, the corresponding term on the rhs of \cref{eq:C2+g} decays quickly for $x\gg 1$, so the asymptotic form of $f_2(x)$ is fully determined from that of $g(x)$.

We end this subsection with a remark about the constant $C_2$. This constant can be in principle determined by going to higher orders and requiring that no singularity appears at $w\to0$; however, the resulting equations appear to be  inconsistent. There may be several reasons for this. First, \cref{eq:Q-dynamics_sm} might cease to be valid at higher orders in $1/\log t$. Another possibility is that the ansatz in \cref{eq:Q_ansatz} breaks down at higher orders; for example, one might have to consider scaling functions of $w\sqrt{t}\log t$ at higher orders in $1/\log t$. At present, we do not know if \cref{eq:Q-dynamics_sm} remains valid at higher orders beyond those considered here. 

\subsubsection{Evolution at short times}
In this subsection, we examine the dynamics of fermionic densities at short times, and argue that, for highly excited initial states, the $k=0$ mode decays quickly. 

We first consider a fully excited initial state. Interestingly, the $k=0$ ($k=\pi$) that is slow (fast) to decay at long times, is the fastest (slowest) to decay in this regime; see the left panel of \cref{fig:rho_initial}. This can be seen directly from \red{Eq.~(5)} of the main text. For a uniform initial state, the last term on the rhs of \red{Eq.~(5b)} vanishes. Substituting for $\rho_k (0)= 1$, we find that fermionic densities initially decay at the rate 
\begin{equation}\label{eq:unit_filling}
    \frac{\rm d}{{\rm d}t}\rho_k \sim \Gamma |e^{ik}+1|^2 \,.
\end{equation}
The dependence on $k$ should be
contrasted against the single-particle decay rate, $\Gamma_k=\Gamma |e^{ik}-1|^2$, introduced in the main text. The origin of their difference can be best understood 
\begin{figure}[t!]
    \begin{center}
    \includegraphics[width=1\linewidth]{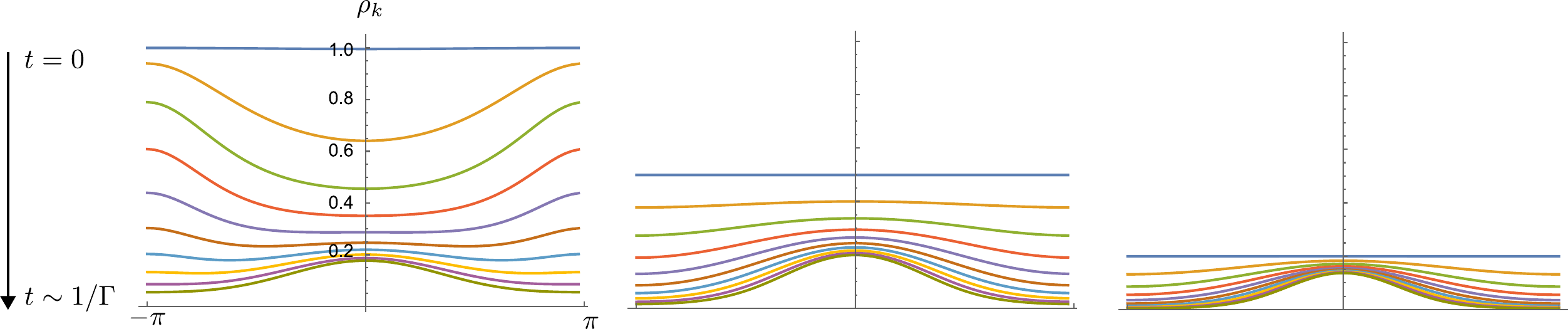}
    \caption{Short-time evolution of $\rho_k(t)$ from \red{Eq.~(5)} of the main text for initial fillings $n(0)=0.2,\,0.5,\,1$ (right, middle, left panels, respectively), at times $\Gamma t=0,\dots,1.25$ in steps $\Gamma \delta t=0.125$. The initial uniform distribution quickly decays to a bell-shaped distribution. Specifically, for $n(0)=1$, the zero mode initially decays more quickly than the other modes and is significantly depleted within a short time, $\Gamma t \sim 1$.}
    \label{fig:rho_initial}
    \end{center}
\end{figure}
from a particle-hole transformation. Starting from a fully excited state, it is natural to make a transformation where $\sigma^z \to \tilde \sigma^z = -\sigma^z$ and $\sigma^{\pm} \to \tilde \sigma^{\mp}$. This is effectively a $\pi$ rotation around the $x$ axis which flips $\sigma^{y,z}$ while keeping $\sigma^x$ intact. The XX Hamiltonian in invariant under this transformation, $\hat H = -2J\sum_j\tilde \sigma^+_j \tilde \sigma^-_{j+1} +{\rm h.c.}$, while the dissipation changes from \cref{eq:newL} to $\hat L_j = \tilde \sigma_{j+1}^- - \tilde \sigma_j^-$. Using the standard Jordan-Wigner mapping to fermions now denoted by $\hat d_j, {\hat d}_j^\dagger$, the Hamiltonian takes the same form in terms of the new fermionic operators, while the dissipator becomes
\begin{equation}
    \hat L_j = \hat {\cal S}(j) ( \hat d_j^\dagger - \hat d_{j+1}^\dagger) \,.
\end{equation}
Writing this expression in momentum space, and ignoring the string operator at low densities ($\langle\hat d_j^\dagger \hat d_j\rangle\ll 1$), 
we can see that these fermions are initially pumped at the rate $P_k = \Gamma |e^{ik}-1|^2$. Transforming back to the original basis, we note that 
\begin{equation}
    \sigma^-_j = {\hat {\cal S}(j)}\hat c_j^\dagger = {\hat {\cal S}^d(j)}\hat d_j  = \tilde\sigma^+_j\,,
\end{equation}
where $\hat {\cal S}^d(j)=\exp(i\pi\sum_{l=1}^j \hat d_l^\dagger \hat d_l)$. 
The above equation is then satisfied upon identifying 
\begin{equation}
    \hat d_j^\dagger = (-1)^{j-1} \hat c_j\,,
\end{equation}
where the overall factor follows from the fact that $\hat {\cal S}^d(j)=(-1)^j \hat {\cal S}(j)$. It thus follows that $\hat d_k^\dagger \sim \hat c_{k+\pi}$ up to an overall phase factor, thus fermions in the original basis are depleted at the rate $P_{k+\pi}$, reproducing \cref{eq:unit_filling}. 

Let us also consider an initial state at half filling. In this case, one can see from \red{Eq.~(5)} of the main text that initially
\begin{equation}\label{eq:half_filling}
     \frac{\rm d}{{\rm d}t}\rho_k \sim \Gamma\,,
\end{equation}
independent of $k$, that is,  different modes initially decay uniformly. 
Finally, for low initial densities, the fermionic densities decay at the rate $\Gamma_k =\Gamma|e^{ik}-1|^2$. We plot the initial decay of fermionic densities at different initial fillings ($n(0)=0.2, 0.5,1$) in \cref{fig:rho_initial}. Interestingly, we find that, regardless of the initial state, the $k=0$ mode's density quickly drops roughly to $\rho_0 \lesssim 0.2$ before the scaling ansatz holds.  

The initial fast decay for highly excited initial states can be understood from the simple approximate treatment presented in the main text where we assumed that $\rho_k(t)$ is roughly uniform in $k$ at short times up to $\Gamma t\sim 1$.
With this approximation together with \red{Eq.~(6)} of the main text, we obtain $\rho_0(t) =1/(\rho_0(0)^{-1}+ 4 \Gamma t)$.
This approximation works particularly well for  an initial state at half filling where different modes initially decay at the same rate (see \cref{eq:half_filling}), and we find an excellent fit to the resulting equation at short times. 
More generally, the linear increase occurs for any initial state (especially if highly excited). In \cref{fig:rho0(t)}, we plot the inverse fermionic density at $k=0$ starting from a fully excited state, clearly showing an initial linear increase in time followed by a sluggish logarithmic growth. 

\begin{figure}[t!]
    \begin{center}
    \includegraphics[width=0.32\linewidth]{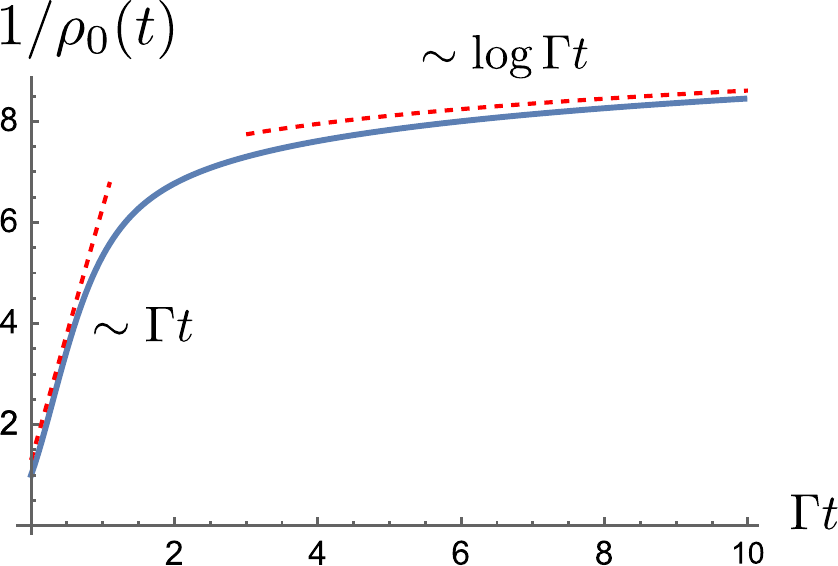}
    \caption{Inverse zero-mode density $1/\rho_0(t)$ as a function of time starting from a fully excited initial state. The short-time linear increase in time crosses over to the logarithmic growth at longer times at around $\Gamma t \sim 1$.}
    \label{fig:rho0(t)}
    \end{center}
\end{figure}

\subsubsection{Universality at long times}
Here, we show that the asymptotic behavior is independent of the initial state. In \cref{fig:initial_n0}, we find that the inverse $\rho_0(t)$ exhibits the same logarithmic scaling down to the same coefficient, i.e., with the same slope in the log-linear plot.

\begin{figure}[h!]
  \centering
  \includegraphics[width=0.3\textwidth]{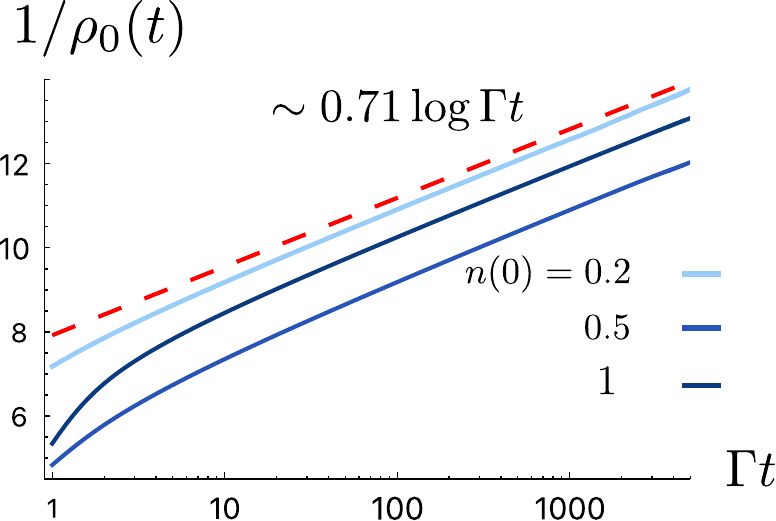}
  \caption{Universality of the late-time decay of the zero mode. Inverse zero-mode density $1/\rho_0(t)$ is plotted vs $\log \Gamma t$ for several initial fillings $n(0)=0.2,0.5,1$. Regardless of the initial state, the decay closely follows the predicted universal scaling, $\rho_0(t)^{-1} \sim (2/\pi)\log \Gamma t$, including the slope.}
  \label{fig:initial_n0}
\end{figure}

\subsubsection{Bosonic momentum distribution}
The fermionic correlation function is given by 
\begin{equation}
    \langle \hat c_j^\dagger c_0\rangle = \int_{-\pi}^\pi \frac{{\rm d}k}{2\pi} \rho_k \cos(kj) = 
    \begin{cases}
    a_0, & j=0 \\
    \frac{1}{2}a_{|j|}, & j\ne 0
    \end{cases}
\end{equation}
where we have used $\rho_k =\rho_{-k}$.

\begin{figure}[t!]
    \begin{center}
    \includegraphics[width=0.9\linewidth]{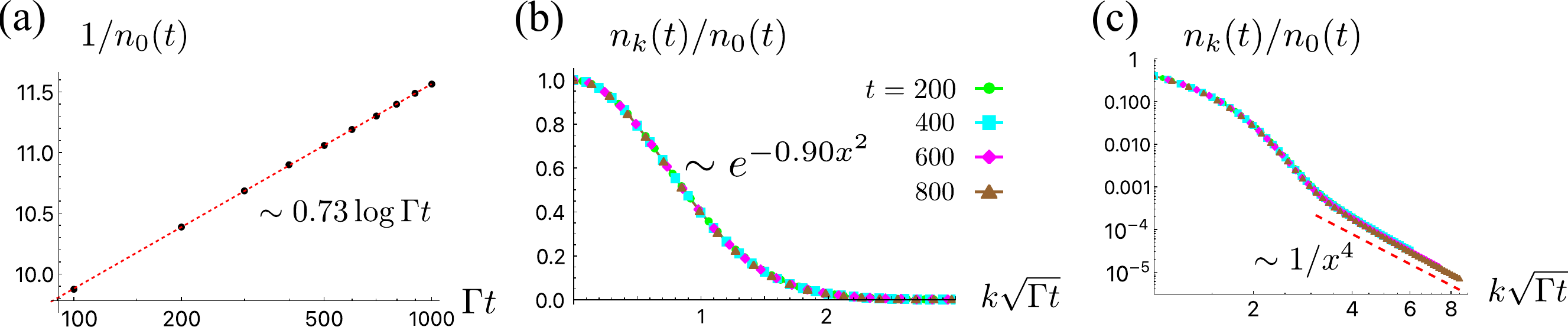}
    \caption{Bosonic momentum distribution $n_k(t)$ defined from spins' (hard-core bosons') correlation function. This exhibits similar characteristics as its fermionic counterpart $\rho_k(t)$: (a) The bosonic zero mode decays in the same universal fashion. (b,c) The normalized bosonic momentum distribution $n_k(t)/n_0(t)$ shows a scaling collapse in $k\sqrt{\Gamma t}$, approximately as a Gaussian distribution for $|k|\sqrt{\Gamma t}  \le 1$, and falling off as  $1/k^4$ at larger momenta.}
    \label{fig:ode_bosonic}
    \end{center}
\end{figure}

The spin correlation function is given by $C(j)\equiv\langle \sigma^+_{j+l} \sigma^-_{l}\rangle$ which can be written in terms of fermions  as 
\begin{equation}
C(j) \equiv \left\langle \hat c_j^\dagger \prod_{l=1}^{j-1}(-1)^{\hat c_l^\dagger \hat c_l}\hat c_0\right\rangle \,.
\end{equation}
This in turn can be computed from  the Toeplitz determinant
\begin{equation}
    C(j)= 2^{j-1} 
\left|
\begin{array}{cccc}
G(1) & G(2) & \cdots & G(j) \\
G(0) & G(1) &  & \vdots \\
\vdots  &   & \ddots & G(2)  \\
G(2-j) & \cdots & G(0) & G(1) 
\end{array}
\right|\,,
\end{equation}
with 
\begin{equation}
    G(j)=\langle \hat c_j^\dagger \hat c_0\rangle -\frac{1}{2}\delta_{j0}\,.
\end{equation}
Viewing spin-$\frac{1}{2}$ particles as hard-core bosons, we can also identify the bosonic  momentum distribution by Fourier transforming the correlation function: 
\begin{equation}
    n_k = \tilde C(k)\,.
\end{equation}
In \cref{fig:ode_bosonic}, we show that $n_k(t)$ shares the same characteristics features with $\rho_k(t)$: (a) $n_0(t)$ shows inverse logarithmic scaling where even the prefactor is approximately the same as that of $\rho_0(t)$; (b) The normalized momentum distribution is a good fit to Gaussian function, $n_k(t)/n_0(t)\approx \exp(-k^2 t)$; and, (c) $n_k(t)\propto 1/k^4$ at large $k$ with the normalized momentum distribution function showing a scaling collapse in $k\sqrt{\Gamma t}$.

\section{MPS simulations}\label{sec:MPS}
In this section we present the matrix product state (MPS) simulations that test the analytical predictions of \cref{sec:GGE} outside the weak-dissipation and thermodynamic limits in which they are derived.  \Cref{subsec:MPS_methods} describes the numerical setup and \cref{subsec:MPS_results} presents the results, comparing the total density, trajectory-averaged entanglement entropy and momentum-resolved fermionic and bosonic observables against \red{Eqs.~(7-8,12)} of the main text. We find the predicted scaling forms with at most factor-of-2 deviations in the fitted coefficients at the finite $\Gamma/J$ and system sizes accessible to MPS.

\subsection{Technical implementation details}\label{subsec:MPS_methods}
We simulate the dissipative dynamics of the XX spin chain defined in \red{Eq.~(2)} of the main text. 
The open-system dynamics is unraveled into stochastically evolving pure MPS using the quantum-trajectory algorithm~\citeSM{paeckel_time-evolution_2019,daley_quantum_2014}, with
all observables obtained by averaging over the ensemble of trajectories.

In practice, we map the XX spin chain to hard-core bosons. This makes the generalization from hard- to soft-core bosons, as in \red{Fig.~(3)} of the main text, a straightforward change in the local Hilbert space dimension.  The  effective Hamiltonian reads
\begin{align}
    \hat H_\mathrm{eff} = -\sum_{j=1}^{L-1} J(\hat b_j^\dagger \hat b_{j+1}+{\rm h.c.})+ \frac{i\Gamma}{2}(\hat b^\dagger_{j+1}-\hat b^\dagger_j)(\hat b_{j+1}-\hat b_j)  + \sum_{j=1}^{L} \frac{U}{2}\hat n_j(\hat n_j-1)\,.
\end{align}
This Hamiltonian follows the conventions for the Bose-Hubbard Hamiltonian; it corresponds to the XX Hamiltonian of \red{Eq.~(2)} of the main text with $J_\mathrm{spin}=1/2$, so the loss rates quoted in figure captions and legends should be multiplied by a factor of 2 when using that convention, $\Gamma/J_\mathrm{spin}=2\,\Gamma/J$. The largest system size considered is $L=120$ and we use open boundary conditions, as convenient for MPS. The initial states are product states obtained by exciting $M$ randomly selected sites among $L$, corresponding to a fixed filling fraction $n(0) = M/L$.

The dynamics is split in two parts. Between jumps we evolve under $\hat H_\mathrm{eff}$ via 2nd-order TEBD~\citeSM{paeckel_time-evolution_2019}: each two-site gate is followed by an SVD truncation to a maximum bond dimension $\chi$, with a total Trotter error per time interval scaling as $\delta t^2$. 
The quantum-jumps use the waiting-time variant: a uniform random threshold $r\in[0,1]$ is drawn, the norm of the state decays under $\hat H_\mathrm{eff}$, and a jump is triggered at the first time step for which $\|\psi\|^2<r$~\citeSM{daley_quantum_2014}. The jump site $j$ is selected with probability proportional to $\langle \hat L_j^\dagger \hat L_j\rangle$, after which the state is renormalized. For convergence, the time step $\delta t$ must be small enough that the probability of two jumps in one step remains negligible, $p_\mathrm{jump}=\delta t \, \Gamma \sum_j \langle \hat L_j^\dagger \hat L_j \rangle \ll 1$. This constrains the computationally accessible system size, initial filling fraction and loss rate.

The effective Hamiltonian $\hat H_\mathrm{eff}$ commutes with the total particle number $\hat N = \sum_j \hat n_j$, so the non-Hermitian evolution between jumps remains within a fixed-$N$ sector. Each jump operator $\hat L^\mathrm{MPS}_j = \hat b_{j+1}-\hat b_j$ maps a state of well-defined particle number to another state of well-defined particle number. The wavefunction therefore retains a definite $\hat N$ throughout the trajectory, and we can exploit the $U(1)$ number-conserving MPS representation in ITensors~\citeSM{fishman_itensor_2020} for more efficient simulations.

Observables are computed from the MPS at runtime. We extract the total density $n\equiv \sum_j \langle \hat n_j\rangle /L$, the bosonic single-particle density matrix $C_{ij}=\langle \hat b^\dagger_i \hat b_j  \rangle $ (chiefly for the hard-core case), and the von Neumann entanglement entropy for an equal bipartition $S_\mathrm{vN}$. 
In order to compare the hard-core MPS results with the analytical results and numerical solutions of the GGE equations, we also compute the fermionic single-particle density matrix, which includes the Jordan-Wigner string:
\begin{equation}\label{eq:Cfermi_mps}
    C_{ij}^\mathrm{fermi} \equiv \langle \hat c^\dagger_i\hat c_j\rangle = \Big\langle \sigma_i^+ \prod_{l=i+1}^{j-1} (1 - 2\hat n_l)\, \sigma^-_j \Big\rangle\,.
\end{equation}
The momentum-space occupation $\rho_k$ is then obtained from the discrete sine transform appropriate for open boundary conditions:
\begin{equation}\label{eq:dst_mps}
    \hat c_k = \sqrt{\frac{2}{L+1}}\sum_{m=1}^{L} \sin\!\left(\frac{m k\pi}{L+1}\right)\hat c_m\,,
\end{equation}
with allowed momenta
\begin{equation}\label{eq:k_allowed}
    k \in \frac{\pi}{L+1}  \{1, \cdots, L\}\,.
\end{equation}
All momentum-space results in \cref{fig:mps_sm_fig2} use the smallest nonzero mode $k_\mathrm{min}=\pi/(L+1)$ as a proxy for the thermodynamic-limit zero mode $\rho_0$.

We ensure convergence of all results in both bond dimension and Trotter time step $\delta t$, checking up to $\chi\leq128$ and $\delta t\geq5\times 10^{-5}$.

\begin{figure}
    \centering
    \includegraphics[width=0.75\linewidth]{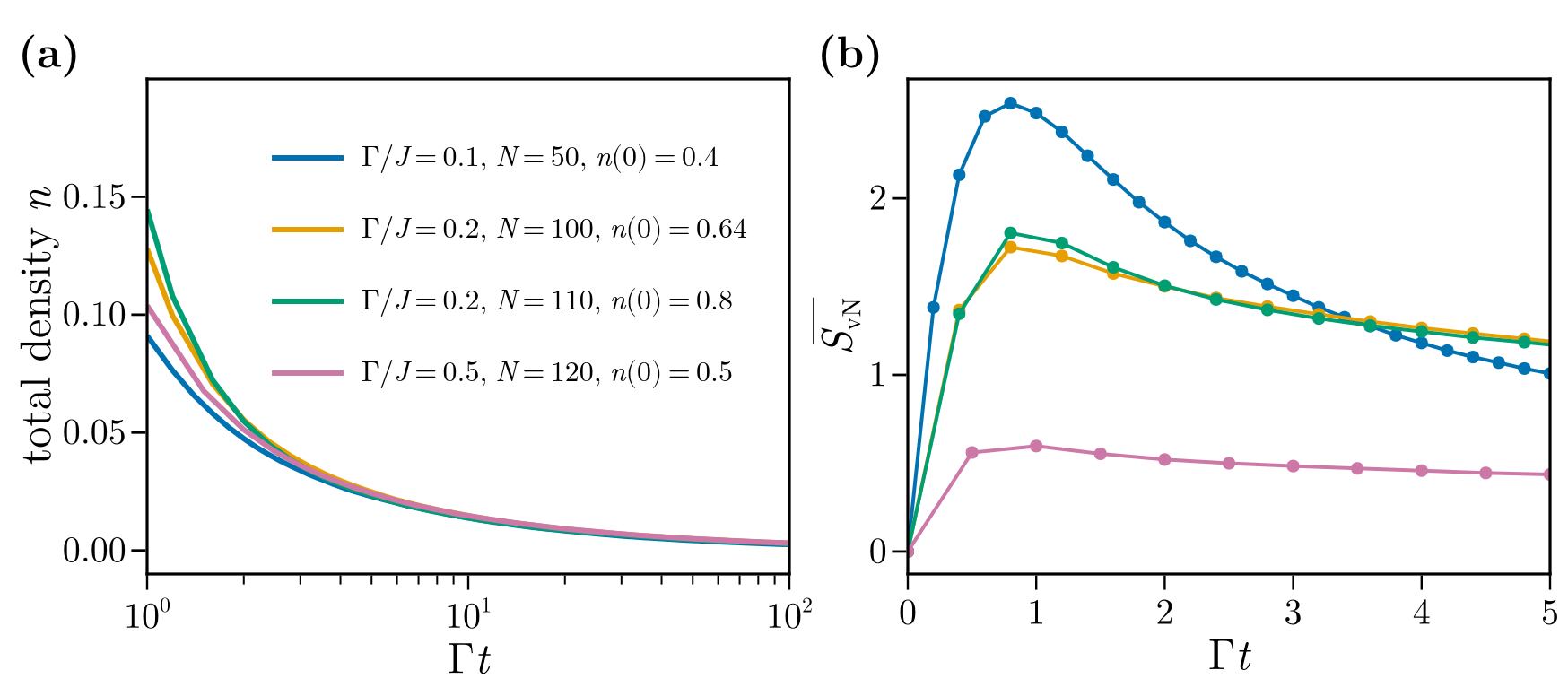}
    \caption{MPS results for the dissipative XX spin chain at various loss rates, system sizes and initial filling fractions (parameters indicated in the legend). (a)~Total population density. (b)~Von Neumann entanglement entropy for an equal bipartition, averaged over quantum trajectories. Statistical error bars are smaller than the line width; $4500 \leq n_\mathrm{traj} \leq 11000$.}
    \label{fig:mps_sm_entropy}
\end{figure}

\subsection{Further numerical results}\label{subsec:MPS_results}

In this section, we provide further numerical results from the MPS simulations. 

\subsubsection{Total density and trajectory-averaged entanglement entropy}
\Cref{fig:mps_sm_entropy}(a) shows the unscaled total density for several system sizes, initial filling fractions and loss rates. The long-time dynamics appear independent of the specific initial conditions and loss rate, and the curves show the same trend at late times. 

For a  bipartition into subsystems $A$ and $B$, the entanglement entropy of a pure state is given by
\begin{equation}
    S_\mathrm{vN}(|\psi\rangle)\equiv-\tr(\rho_A\log\rho_A) = -\tr(\rho_B\log\rho_B)\,, \qquad \rho_A = \tr_B \ketbra{\psi}\,.
\end{equation}
Along each trajectory $\ket{\psi_\alpha}$, we evaluate $S_\mathrm{vN}(\ket{\psi_\alpha})$ for the half-system bipartition and average over the ensemble to obtain $\overline{S_{\rm vN}}$. The resulting trajectory-averaged entanglement is \emph{not} an entanglement measure of the open-system density matrix; it instead tracks the bond-dimension cost of the MPS+QT simulation. \Cref{fig:mps_sm_entropy}(b) shows the average entropy up to $\Gamma t=5$: the entropy rises from zero for the initial product state, peaks around $\Gamma t\sim 1$ (roughly marking the onset of the scaling regime), and then slowly decreases as the system approaches a trivial steady state.

\subsubsection{Fermionic momentum space observables}

\begin{figure}
    \centering
    \includegraphics[width=0.99\linewidth]{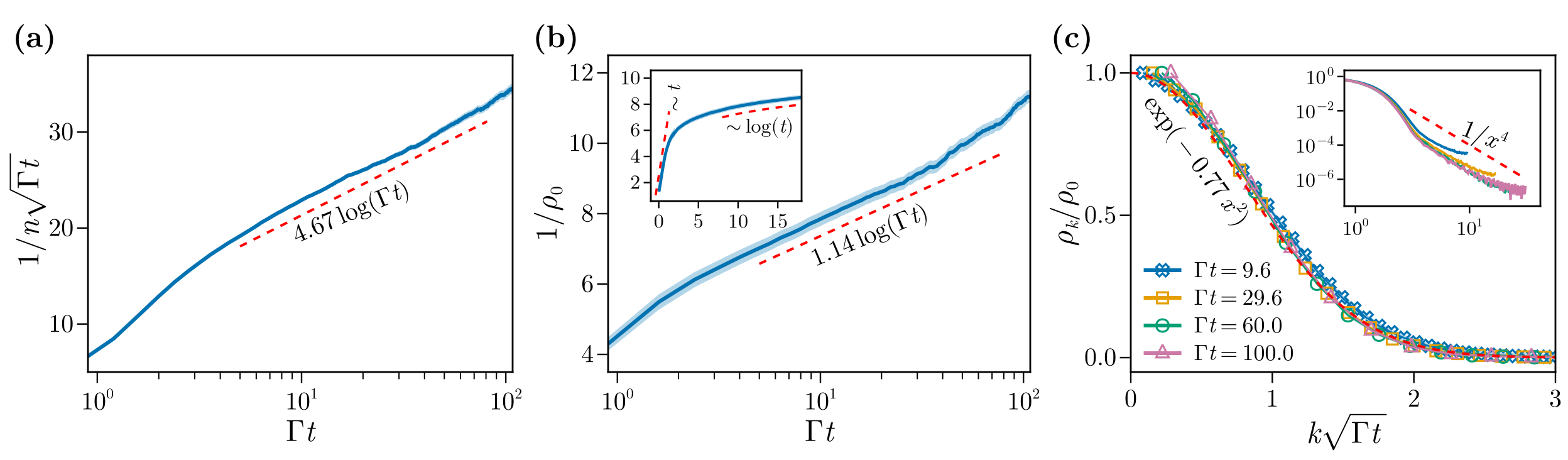}
    \caption{MPS results for the fermionic observables, tested against the analytical predictions of \red{Eqs.~(7,8,12)}. (a)~Inverse rescaled total density $1/(n\sqrt{\Gamma t})$ vs.\ $\log(\Gamma t)$. (b)~Inverse population of the smallest quasi-momentum mode $k_{\min}=\pi/(L+1)$, used as a proxy for the zero mode $\rho_0$, vs.\ $\log(\Gamma t)$; inset: crossover between the early-time linear and the late-time logarithmic regimes. (c)~Normalized momentum distribution $\rho_k/\rho_0$ vs.\ $k\sqrt{\Gamma t}$ at several times; inset: log-log plot showing the crossover to an algebraic $\sim(k\sqrt{\Gamma t})^{-4}$ tail. Dashed red lines are linear or Gaussian fits, as appropriate. Error bars on the momentum-resolved quantities in panels~(b,c) are estimated by treating each trajectory's full correlation matrix as a single correlated sample, which yields a conservative and nearly time-independent bound rather than a faithful trajectory-to-trajectory dispersion. Parameters: $L=110$, $n(0)=0.8$, $\Gamma/J=0.2$, $\chi=64$, $\delta t=0.01\,J^{-1}$, $n_\mathrm{traj}=12880$.}
\label{fig:mps_sm_fig2}
\end{figure}

\Cref{fig:mps_sm_fig2} tests the analytical predictions for the fermionic observables (after JW), \red{Eqs.~(7,12)}, on a finite open spin chain of length $L=110$.
The rescaled total density $(n\sqrt{\Gamma t})^{-1}$ is shown in \cref{fig:mps_sm_fig2}(a), where the straight line gives the logarithmic decay predicted by \red{Eq.~(12)} of the main text via $n=\int\frac{{\rm d}k}{2\pi}\,\rho_k$. The analytical prediction for the inverse rescaled total density is $(n\sqrt{\Gamma t})^{-1}\sim a\log(\Gamma t)$ with prefactor $a=4/\sqrt{\pi}\approx2.257$; the numerical fit [\cref{fig:mps_sm_fig2}(a)] yields $a_{\rm MPS}\approx 4.67$, roughly a factor of $2$ larger.
The inverse zero-mode population is predicted to scale as $\rho_0^{-1}\sim c\log(\Gamma t)$ with $c=2/\pi\approx0.637$ [\red{Eq.~(8)}]; the numerical fit [\cref{fig:mps_sm_fig2}(b)] gives $c_{\rm MPS}\approx 1.14$, again roughly a factor of $2$ larger.
At small rescaled momenta ($k\sqrt{\Gamma t}\lesssim 3$), the curves collapse onto a Gaussian $\rho_k/\rho_0 \simeq \exp(-\alpha\, k^{2}\,\Gamma t)$ [\cref{fig:mps_sm_fig2}(c)], consistent with \red{Eq.~(12)}, but with a fitted exponent $\alpha = 0.77$ rather than unity. The apparent rightward drift of the leftmost data point in panel~(c) as $\Gamma t$ grows is a finite-size artifact of using open boundaries: the smallest accessible momentum is fixed at $k_{\min}=\pi/(L+1)>0$, so the rescaled abscissa $k_{\min}\sqrt{\Gamma t}$ increases with time while the ordinate itself is $\rho_{k_{\min}}/\rho_0\equiv 1$ by construction. The inset confirms the crossover to the predicted algebraic $\sim 1/(k\sqrt{\Gamma t})^{4}$ tail at large momenta, in agreement with the $|k|\sqrt{\Gamma t}\gg 1$ branch of \red{Eq.~(12)}.

The MPS simulations confirm the predicted $\sim \log(\Gamma t)$ scaling of total density $1/(n\sqrt{\Gamma t})$ and zero mode $1/\rho_0$, but the numerically extracted prefactors differ from the analytical values by a factor of almost 2.
Several sources may contribute to this discrepancy: (i)~the analytical prefactors are derived in the thermodynamic limit with periodic boundary conditions, whereas the simulations uses an open chain of finite size $L=110$; (ii)~the weak-dissipation approximation underlying the GGE treatment may receive corrections at the finite ratio $\Gamma/J = 0.4$ used here; and (iii)~the asymptotic regime $\Gamma t \gg 1$ where subleading $1/\log(\Gamma t)^2$ corrections become negligible may not be fully reached within the accessible simulation times.

Finally, the inset of \cref{fig:mps_sm_fig2}(b) shows the early-time linear growth of $1/\rho_0$ 
giving way to the slower logarithmic increase at late times, consistent with the two-regime picture of the previous section.

\subsubsection{Hard-core boson momentum observables}
\Cref{fig:mps_sm_fig2_bosonic} shows the analog of \cref{fig:mps_sm_fig2}(b,c) for the hard-core boson (spin) observables, i.e.\ without the Jordan-Wigner string. The trends mirror the fermionic case with similar fitted coefficients. The bosonic MPS data may also be compared with the GGE results of \cref{fig:ode_bosonic}; we again find agreement within at most a factor of 2.

\begin{figure}
    \centering
    \includegraphics[width=0.7\linewidth]{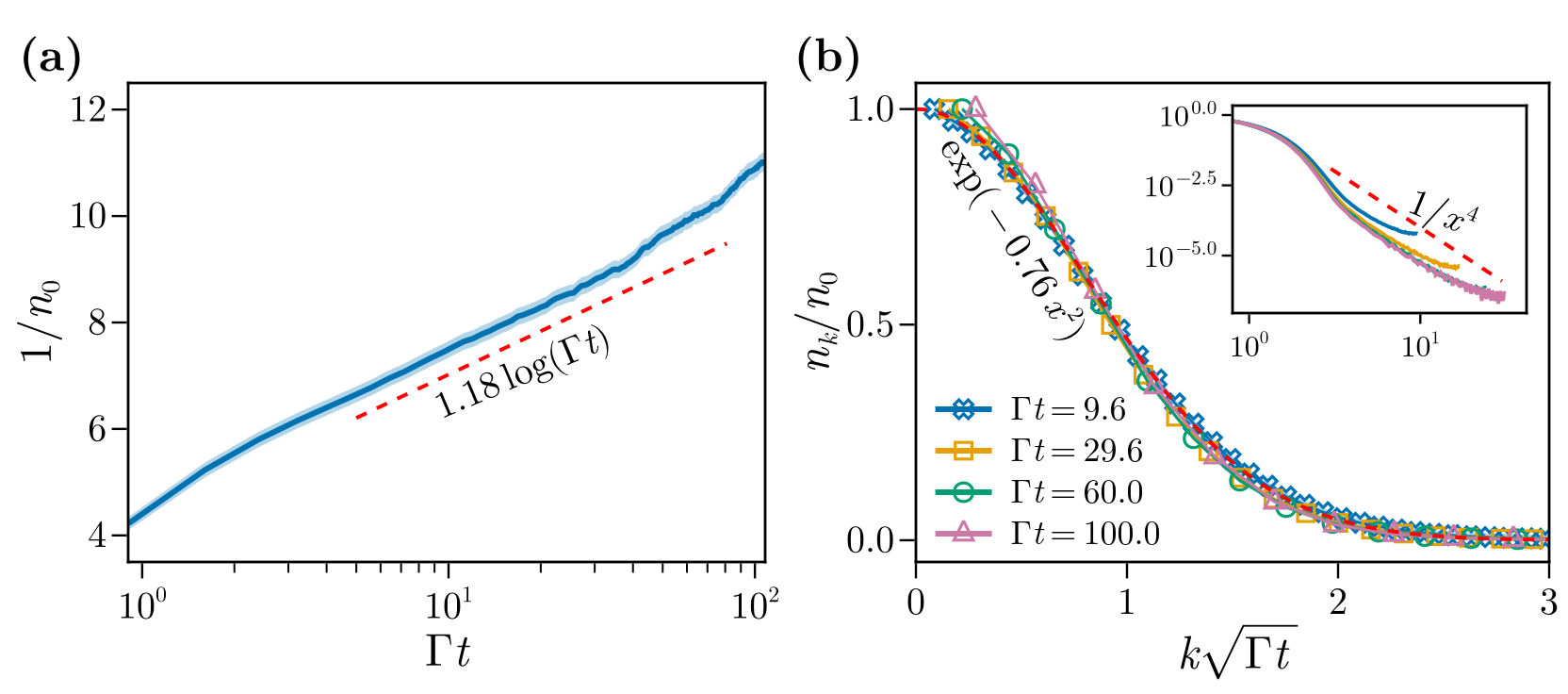}
    \caption{MPS results for the hard-core boson (spin) observables.
    (a)~Inverse population of the smallest quasi-momentum mode $k_{\min}=\pi/(L+1)$, used as a proxy for the zero mode $n_0$, vs.\ $\log(\Gamma t)$. (b)~Normalized momentum distribution $n_k/n_0$ vs.\ $k\sqrt{\Gamma t}$ at several times; inset: log-log plot showing the crossover to an algebraic $\sim(k\sqrt{\Gamma t})^{-4}$ tail. Dashed red lines indicate linear or Gaussian fits, as appropriate. Parameters: same as \cref{fig:mps_sm_fig2}.}
\label{fig:mps_sm_fig2_bosonic}
\end{figure}

\subsubsection{Comparison with the GGE at short times}
\begin{figure}
  \centering
  \includegraphics[width=0.8\textwidth]{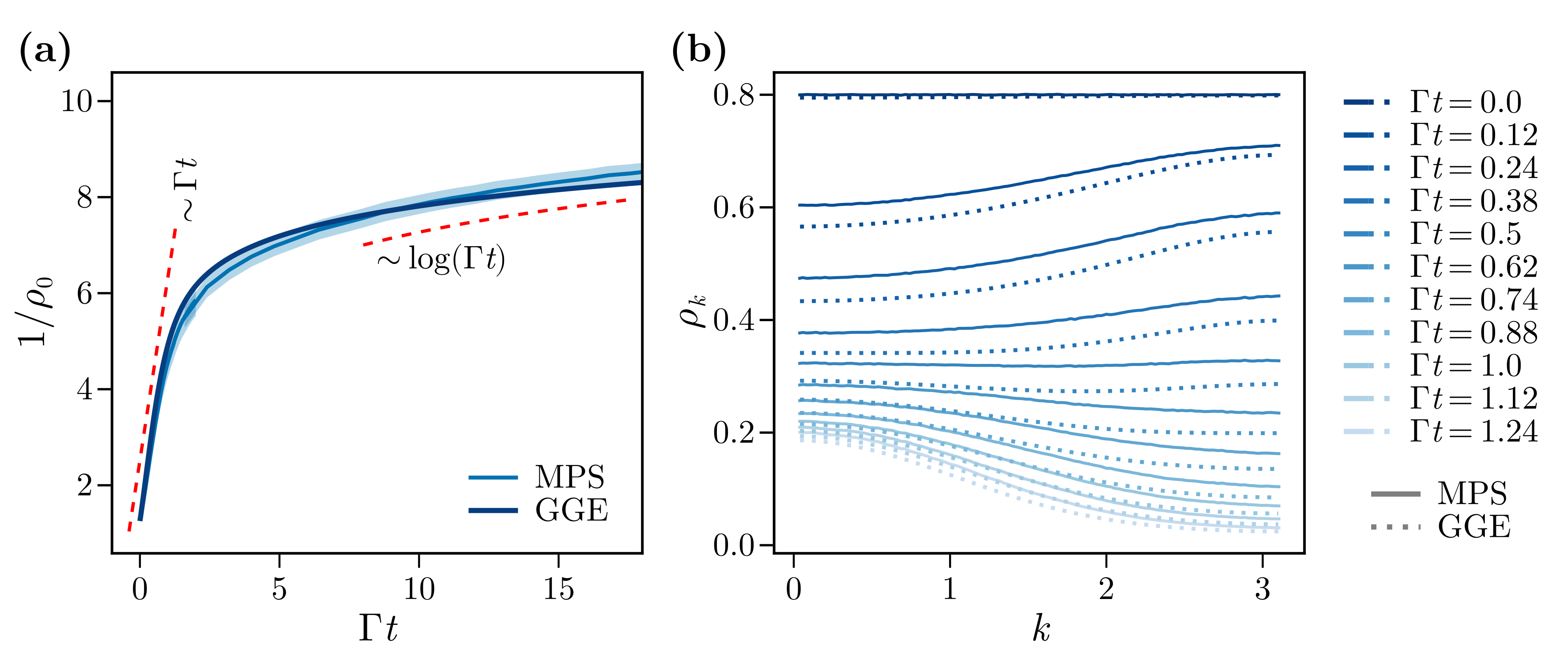}
  \caption{MPS results compared with the solution of the GGE results.
  (a)~Inverse population of the smallest quasi-momentum mode $k_{\min}=\pi/(L+1)$, used as a proxy for the zero mode $\rho_0$, at short times, showing the crossover from $\rho_0\sim 1/t$ to $\rho_0\sim 1/\log(\Gamma t)$ around $\Gamma t\sim 1$. (b)~Momentum distribution $\rho_k$ at several times up to $\Gamma t \sim 1$. Parameters:  same as \cref{fig:mps_sm_fig2}.}
  \label{fig:rhok_compare_mps_ode}
\end{figure}

\Cref{fig:rhok_compare_mps_ode} compares the momentum distribution $\rho_k$ from MPS against the GGE solution with the initial filling $n(0)=0.8$. Panel~(a) shows that both methods reproduce the $\rho_0\sim 1/\Gamma t \to 1/\log(\Gamma t)$ crossover around $\Gamma\tau\sim 1$, and panel~(b) shows good quantitative agreement of the full $\rho_k$ at short times. This agreement is remarkable given the different settings: a finite open chain with $\Gamma/J= 0.2$ versus the GGE solution in the thermodynamic limit and in the limit $\Gamma /J\ll 1$.

\null\vfill\clearpage
\bibliographystyleSM{apsrev4-2}
\bibliographySM{refs.bib}

\fi

\end{document}